\documentclass[%
 reprint,
 amsmath,amssymb,
 aps,
]{revtex4-2}

\usepackage{graphicx}
\usepackage{dcolumn}
\usepackage{bm}
\usepackage{comment}
\usepackage{amsmath}
\usepackage{tikz}
\usepackage{amsmath}
\usepackage{caption}
\usepackage{subcaption}
\usepackage[dvipsnames]{xcolor}
\definecolor{tan}{RGB}{187, 102, 34}
\definecolor{blue}{RGB}{22, 64, 77}
\definecolor{teal}{RGB}{73, 125, 116}

\usepackage[pagebackref=false,
  hyperindex=true,
  citecolor=blue,
  colorlinks=true,
  linkcolor=blue,
urlcolor=blue]{hyperref}
\usepackage{pifont}    
\usepackage{xcolor}

\newif\ifshowcomments
\showcommentstrue 

\begin{document}

\preprint{APS/123-QED}

\title{Effective-one-body modelling of eccentric supermassive black hole binaries for Pulsar Timing Array.}

\author{Sara Manzini}
\email{manzini@apc.in2p3.fr}

\author{Stanislav Babak}
\email{stas@apc.in2p3.fr}

\affiliation{Université Paris Cité, CNRS, Astroparticule et Cosmologie, F-75013 Paris, France}

\date{\today}

\begin{abstract}
Pulsar Timing Arrays (PTAs) observations will detect gravitational waves (GWs) from the early inspiral phase of supermassive black hole binaries (SMBHBs) with orbital periods of weeks to years.  Current PTA analyses generally assume circular binaries; however, dynamical interactions with the surrounding environment can prevent complete circularisation, allowing SMBHBs to retain appreciable eccentricities.
In this work, we present a gravitational waveform model for eccentric binaries based on the Effective-One-Body (EOB) formalism, designed for continuous GW searches in PTA data. The model is accurate up to the second post-Newtonian (2PN) order for the conservative dynamics and up to post-leading order for the radiation-reaction terms. We provide both a numerically precise and a computationally efficient approximate implementation and evaluate the latter’s accuracy against the full model over a broad range of eccentricities and initial orbital frequencies.  Our results show that a substantial region of the parameter space exhibits pronounced orbital evolution, much stronger than in the circular case. We demonstrate the rich harmonic structure of timing residuals induced by eccentric GWs.
Properly characterising eccentric binaries is an essential step toward detecting GWs in PTA data and interpreting the results, ultimately improving our understanding of the supermassive black hole population in the local Universe.
\end{abstract}

\maketitle

\section{\label{sec:intro}Introduction}
Pulsar Timing Array (PTA) collaborations conduct decades-long timing of millisecond pulsars, aiming to detect gravitational waves (GWs) in the nano-Hertz frequency band. The most anticipated sources in this band are inspiralling supermassive black hole binaries in the local Universe \cite{10.1111/j.1365-2966.2008.13682.x, Perrodin_2018}. PTAs are particularly sensitive to systems hosting SMBHs with masses exceeding $10^7\, M_{\odot}$, typically located in the nuclei of massive galaxies. Galaxies and their central SMBHs evolve within the hierarchical structure formation paradigm, where SMBHs grow through gas and dark matter accretion, as well as through successive merger events~\cite{Volonteri_2003, 2019A&ARv..27....5B}. During a merger, galaxies become gravitationally bound on highly eccentric orbits, and this initial eccentricity is inherited by the nascent SMBHB~\cite{Sesana_2010}. The hardening of the bound SMBH pair is primarily driven by dynamical friction against gas, stars, and dark matter, which brings the binary to parsec-scale separations~\cite{Li_2020}.
Numerical simulations indicate that, at this stage, SMBHBs may still retain significant eccentricities, particularly when the interaction is dominated by a gaseous disk that confines the secondary SMBH outside the stellar bulge’s sphere of influence~\cite{Dotti_2007}.
At parsec separations, dynamical friction becomes inefficient, while gravitational wave emission only starts to dominate at separations of order $0.01\,\mathrm{pc}$. One proposed mechanism to bridge this so-called “final parsec problem”~\cite{2003ApJ...596..860M} is through three-body interactions with surrounding stars. In this scenario, the hard binary loses orbital energy by capturing stars passing within a distance comparable to binary separation and ejecting them at high velocities~\cite{Sesana_2006}. Another mechanism involves perturbations from a third SMBH introduced during a subsequent galactic merger. Such a perturber can trigger Kozai–Lidov oscillations, driving large eccentricities in otherwise stalled binaries. This process can produce highly eccentric SMBHBs that radiate in the PTA frequency band.

In 2023, the European PTA (EPTA \cite{10.1093/mnras/stw483}), the North American Nanohertz Observatory for Gravitational Waves (NANOGrav \cite{Demorest:2012bv}), the Parkes PTA (PPTA \cite{Manchester}), and the Chinese PTA (CPTA \cite{Xu_2023}) collaborations jointly reported evidence for an emerging gravitational wave (GW) signal in the nanohertz band \cite{EPTA2023c, Agazie_2023, Reardon_2023, Xu_2023}. The signal is consistent with a stochastic gravitational wave background (GWB) generated by a population of supermassive black hole binaries (SMBHBs), though its precise nature remains under investigation. Most analyses performed by the PTA community assume circular binaries. However, the spectral shape of the GWB produced by a population of eccentric binaries is suppressed at low frequencies compared to the idealised power-law spectrum used to describe circular binaries~\cite{Chen:2016zyo}. Moreover, a population of strongly eccentric binaries could violate the assumptions of isotropy, Gaussianity, and stationarity commonly adopted in GWB analyses~\cite{PhysRevD.109.123010, falxa2024eccentricbinariesnonstationarygravitational}. Accurate modelling of eccentric binaries is therefore essential for identifying such features, which may help distinguish an astrophysical GWB from one of cosmological origin.

PTA observations are primarily sensitive to the early, long-lasting inspiral phase of individual, slowly evolving SMBHBs. These systems are observed far from coalescence, with orbital periods between a few months and a few years. A significant fraction of these SMBHBs are expected to retain measurable eccentricities, particularly in scenarios involving the ``perturber'' mechanism described above. In alternative formation channels, non-negligible dynamical interactions between SMBHBs and their surrounding environment can inhibit efficient circularisation through gravitational wave emission~\cite{Taylor_2016}. 

Circular binaries emit gravitational waves predominantly at twice the orbital frequency; higher modes are strongly suppressed by post-Newtonian factor $v/c$ -- ratio of orbital velocity to the speed of light.  In contrast, eccentric binaries radiate across a broad spectrum of harmonics, corresponding to integer multiples of the (mean) orbital frequency with the side-bands due to the orbital precession. Such systems also evolve more rapidly than circular binaries with the same mean frequency. Moreover, SMBHBs experience a non-negligible relativistic precession of the periapsis, which must be properly accounted for. Therefore, an accurate template for eccentric binaries is crucial to improve the detection prospects of individually resolvable systems and to reduce biases in parameter inference.

An initial effort to model the waveform of eccentric binaries was carried out by \cite{Taylor_2016} and \cite{PhysRevD.92.063010}, using a phenomenological approach introduced in \cite{PhysRev.131.435} and \cite{Barack:2003fp}. Searches for eccentric binaries in PTA data were performed by \cite{2004ApJ...606..799J}, \cite{10.1093/mnras/stv381}, \cite{2024A&A...690A.118E}, and more recently by \cite{NANOGrav:2023wsz}. The search presented in \cite{NANOGrav:2023wsz} is particularly relevant to our study, as it is based on the model developed in a series of works \cite{PhysRevD.96.044011, PhysRevD.101.043022, Susobhanan:2022nzv}. This model relies on the quasi-Keplerian formalism in ADM coordinates and describes orbital evolution up to leading order in eccentricity and frequency, including periapsis precession at post-leading order. Moreover, it is converted into PTA timing residuals by incorporating the appropriate response function.
Another eccentric model for the gravitational wave signal was recently proposed in \cite{PhysRevD.111.084052}. While this model is the most complete to date - accounting also for spin-orbit precession - it is formulated in the frequency domain and is therefore not directly suitable for PTA data analysis, where observations are unevenly sampled.

In this paper, we introduce a new template based on the Effective One Body (EOB) approach~\cite{Buonanno:1998gg}. In particular, our derivations build upon~\cite{Hinderer_2017}, where the conservative dynamics are described in terms of the true anomaly $\xi$ and the azimuthal orbital phase $\phi$. Due to orbital (periapsis) precession, the corresponding radial and azimuthal orbital frequencies differ already at 1PN order.

We define the orbital eccentricity uniquely via the turning points of the radial motion, without the need for auxiliary eccentricities commonly used in post-Keplerian descriptions \cite{PhysRevD.101.043022, PhysRevD.70.104011}. This formalism provides an intuitive representation of binary evolution, described in Section~\ref{sec:dynamics}, where we present the conservative dynamics ($\phi(t)$ and $\xi(t)$ in Section~\ref{sec:conservative}) and introduce dissipative terms associated with gravitational wave emission, leading to the evolution of orbital frequencies and eccentricity (Section~\ref{sec:dissipative}).
Starting from the formulation in~\cite{Hinderer_2017}, we perform a change of variables to adapt the dynamical equations to the astrophysical population of SMBHBs in PTA band. We further extend the evolution to 2PN order for the conservative part and to next-to-leading order for the dissipative terms.

In Section~\ref{sec:waveform}, we compute gravitational wave strain polarizations $h_{+,\times}$  which could also be used to model GWs from eccentric Galactic binaries in the LISA band. In Section~\ref{sec:residuals}, we derive the timing residuals relevant for PTA data analysis and present the ready-to-use template in time and frequency
domains. 

Section ~\ref{sec:results} is split into two parts:
\begin{itemize}
    \item In Section~\ref{sec:solution_ODEs}, we introduce an approximate solution (~\ref{sec:dynamics_approx}) to orbital dynamics to improve computational efficiency, while quantifying accuracy compared to the full numerical solution in Section~\ref{sec:dynamics_num}. 
    \item In Section~\ref{sec:fourier_decomposition}, we describe the GW induced residuals in frequency domain using the stationary phase approximation.
\end{itemize}

Eventually,  Section~\ref{sec:conclusion} summarize the content of this paper.

The Appendix~\ref{sec:ADM} provides the mapping between the coordinate system used in our EOB-based formulation and that adopted in other eccentric waveform models~\cite{PhysRevD.96.044011, PhysRevD.101.043022, Susobhanan:2022nzv}, while Appendix~\ref{sec:fourier} contains explicit calculation of terms used in the frequency domain expression of residuals.

\section{\label{sec:dynamics}Dynamics}

The Effective One Body (EOB) formalism~\cite{Buonanno:1998gg, PhysRevD.62.064015, PhysRevD.105.044035, PhysRevD.62.084011, Damour_2001}  reformulates the relativistic two-body problem into an equivalent description of a single test particle of reduced mass $\mu = m_1 m_2 / M$, where $M = m_1 + m_2$ is the total mass, moving in an effective external spacetime. This effective metric represents a static, spherically symmetric deformation of the Schwarzschild geometry, with the symmetric mass ratio $\nu = \mu / M$ acting as the deformation parameter.

For non-spinning binaries, the effective metric takes the form
\begin{equation}\label{eq:metric}
ds_{\text{eff}}^2 = -Adt^2 + \frac{dr^2}{A D} + r^2 d\phi^2,
\end{equation}
where the potentials $A \to 1 - 2M/r$ and $D \to 1$ reduce to the Schwarzschild case in the test-particle limit $\nu \to 0$.

The Effective One Body (EOB) Hamiltonian \cite{Buonanno:1998gg} is given by
\begin{equation}\label{eq:Heob}
H_{\text{EOB}} = M \sqrt{1 + 2\nu \left( \frac{H_{\text{eff}}}{\mu} - 1 \right)},
\end{equation}
with
\begin{equation}\label{eq:Heff2}
H_{\text{eff}}^2 = A \left[\mu^2 + \frac{P_{\phi}^2}{r^2} + A D P_r^2 + \frac{Q_4(r) M^2 P_r^4}{\mu^2 r^2} + \mathcal{O}(P_r^6) \right],
\end{equation}
where $P_r$ and $P_{\phi}$ denote the canonical radial and azimuthal momenta.
The function
$Q_4 = 2(4 - 3\nu)\nu + \mathcal{O}(r^{-1})$ accounts for non-geodesic corrections that enter at third post-Newtonian (3PN) order~\cite{Hinderer_2017, Damour_2001, PhysRevD.62.084011}.

The orbital dynamics are governed by the Hamiltonian equations:
\begin{eqnarray}\label{eq:EOMs}
\dot{r} &= \frac{\partial H_{\text{EOB}}}{\partial P_r}, \label{subeqn:EOMs-r}\\
\dot{\phi} &= \frac{\partial H_{\text{EOB}}}{\partial P_{\phi}}, \label{subeqn:EOMs-phi} \\ 
\dot{P}_r &= - \frac{\partial H_{\text{EOB}}}{\partial r} + \mathcal{F}, \label{subeqn:EOMs-Pr} \\
\dot{P}_{\phi} &= - \frac{\partial H_{\text{EOB}}}{\partial \phi} + \mathcal{G}, \label{subeqn:EOMs-Pphi}
\end{eqnarray}
where the first term in each equation represents the conservative part of the dynamics, and $\mathcal{F}$ and $\mathcal{G}$ are the radial and azimuthal components of the radiation-reaction forces arising from gravitational-wave emission. These dissipative terms lead to the loss of orbital energy and angular momentum~\cite{Gamboa:2024imd, PhysRevD.86.124012, PhysRevD.67.064028, PhysRevD.62.044024, PhysRevD.78.024009}.

\begin{figure*}[t]
    \centering
    \begin{subfigure}{0.45\textwidth}
        \centering
        \begin{tikzpicture}[scale=1.5, every node/.style={font=\Large}]
    \def\a{2}           
    \def\b{1.5}         
    \def\c{1.32}        
    \def\E{60}          
    \def\omegaval{30}      
    
    \def\nu{95}         
    
    \coordinate (O) at (0,0);           
    \coordinate (F1) at (\c,0);        
    \coordinate (F2) at (-\c,0);         
    
    \coordinate (P) at ({{\a*cos(\E)}, {\b*sin(\E)}});
    
    \coordinate (Q) at ({{\a*cos(\E)}, {\a*sin(\E)}});
    
    \def\M{40}          
    \coordinate (M) at ({{\a*cos(\M)}, {\a*sin(\M)}});
    
    \draw[thin, black, dashed] (O) circle (\a);
    
    \draw[thick, black] (O) ellipse ({\a} and {\b});
    
    \fill[black] (F1) circle (0.08);
    \node[black, below = 0.13cm] at (F1) {$M$};
    
    \fill[black] (O) circle (0.03);
    \node[below left] at (O) {$O$};
    
    \fill[black] (P) circle (0.05);
    \node[black] at (1.2,1.4) {$\mu$};
    
    \draw[thick, black] (-\a-0.5, 0) -- (\a+0.5, 0);
    
    \draw[thick, black, dashed] (Q) -- (P);
    \draw[thick, black, dashed] ({\a*cos(\E)},0) -- (Q);
    
    \draw[ultra thick, teal] (O) -- (Q);
    \draw[ultra thick, teal, ->] (0.85,0) arc (0:\E:0.85);
    \node[teal] at (0.7, 0.8) {$u$};
    
    \draw[ultra thick, tan] (O) -- (M);
    \draw[ultra thick, tan, ->] (0.3,0) arc (0:\M:0.3);
    \node[tan] at (0.6, 0.2) {$\psi_r$};
    
    \draw[ultra thick, blue] (F1) -- (P);
    \draw[ultra thick, blue, ->] ({\c + 0.295}, 0) arc (\c + 0.1:\nu:0.32);
    \node[blue] at ({\c + 0.2}, 0.43) {$\xi$};

    \coordinate (Peri) at (\a, 0);
    \coordinate (Apo) at (-\a, 0);
    
\end{tikzpicture}
        \caption{Diagram representing the eccentric binary dynamics, M is the total mass of the binary, $\mu$ is the reduced mass and $r$ is the radial separation. The phases are the mean anomaly $\psi_r$, the eccentric anomaly $u$ and the true anomaly $\xi$}
    \end{subfigure}
    \hfill
    \begin{subfigure}{0.45\textwidth}
        \centering
        \begin{tikzpicture}[scale=1.5, every node/.style={font=\Large}]
    \def\a{2}           
    \def\b{1.5}         
    \def\c{1.32}        
    \def\E{60}          
    \def\rotation{20}   
    
    \def\nu{95}         
    
    \coordinate (F1) at (\c,0);
    
    \coordinate (O) at ({\c - \c*cos(\rotation)}, {-\c*sin(\rotation)});
    
    \coordinate (F2) at ({\c - 2*\c*cos(\rotation)}, {-2*\c*sin(\rotation)});
    
    \coordinate (P) at ({{\c + (\a*cos(\E) - \c)*cos(\rotation) - \b*sin(\E)*sin(\rotation)}, {(\a*cos(\E) - \c)*sin(\rotation) + \b*sin(\E)*cos(\rotation)}});
    
    \begin{scope}[shift={(F1)}, rotate=\rotation, shift={(-\c,0)}]
        \draw[thick, black] (0,0) ellipse ({\a} and {\b});
    \end{scope}
    \draw[thin, black, dashed] (0,0) ellipse ({\a} and {\b});
    
    \fill[black] (F1) circle (0.08);
    \node[black, below = 0.13cm] at (F1) {$M$};
    
    \fill[black] (O) circle (0.03);
    \node[below left] at (0.2,-0.5) {$O$};
    
    \fill[black] (P) circle (0.05);
    \node[black] at ([shift={(-0.2,-0.2)}]P) {$\mu$};
    
    \coordinate (PeriRot) at ({\c + ((\a + 0.75) - \c)*cos(\rotation)}, {((\a + 0.75) - \c)*sin(\rotation)});
    \coordinate (ApoRot) at ({\c + (-(\a + 0.75) - \c)*cos(\rotation)}, {(-(\a + 0.75) - \c)*sin(\rotation)});
    \draw[thick, black] (ApoRot) -- (PeriRot);
        
    \draw[ultra thick, Orchid] (F1) -- (0,2);
    \draw[ultra thick, Orchid, ->] (2.4,0) arc (0:100:2);
    \node[ultra thick, Orchid] at (2, 1.5) {$\phi$};

    \draw[ultra thick, blue, dashed] (F1) -- (P);
    \draw[ultra thick, blue, ->] ({\c + 0.285}, 0.1) arc (0:{atan2((\a*cos(\E) - \c)*sin(\rotation) + \b*sin(\E)*cos(\rotation), (\a*cos(\E) - \c)*cos(\rotation) - \b*sin(\E)*sin(\rotation))}:0.32);
    \node[ultra thick, blue] at (1.1,0.7) {$\xi$};

    \draw[ultra thick, olive] (F1) -- (PeriRot);
    \draw[ultra thick, olive, ->] ({\c + 1.45}, 0) arc (1:\rotation:1.4);
    \node[ultra thick, olive] at ({\c +1.6}, {0.2}) {$\gamma$ };
    

    \draw[thin, black, dashed] (-2.75, 0) -- (2.75, 0);
    
\end{tikzpicture}
        \caption{In this diagram we represent a time evolution of diagram (a). $\phi$ is the total azimuthal phase, its time derivative $\dot{\phi}$ differs from true anomaly derivative $\dot{\xi}$ at 1PN order, leading to the periastron precession $\gamma$.}
    \end{subfigure}
    \caption{Two diagrams representing the eccentric binary properties and phases.}\label{fig:orbit}
\end{figure*}
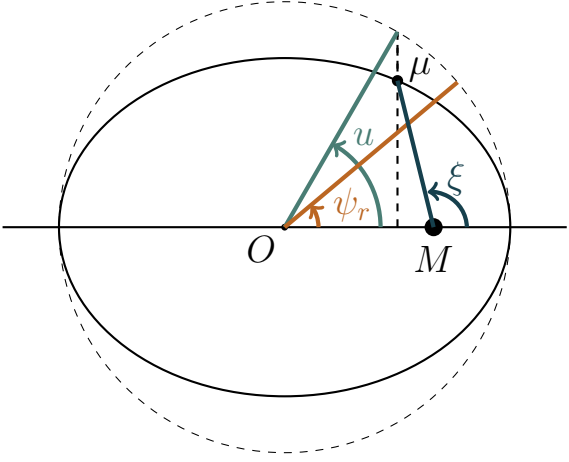
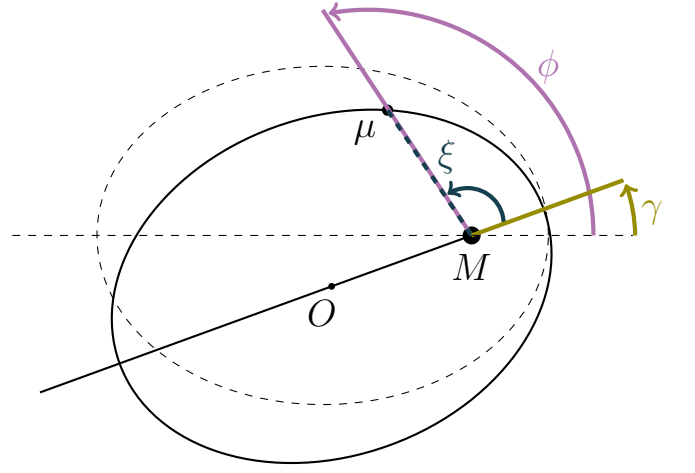


\subsection{\label{sec:conservative}Conservative Dynamics}
To describe the conservative dynamics of the binary, it is useful to express the radial coordinate in terms of the true anomaly $\xi$:
\begin{equation}\label{eq:rfuncxi}
    r = \frac{pM}{1 + e \cos \xi},
\end{equation}
where $p$ is the semilatus rectum and $e$ is the eccentricity of the binary. The latter is uniquely defined by the turning points of the radial motion $r_{1,2} = \frac{pM}{1\mp e} $ which occur at $\xi = 0, \pi$.

The phase $\dot{\xi}$ is continuous and its introduction eliminates the need to change the sign of $r$ after each turning point:
\begin{equation}\label{eq:xidot}
    \dot{r} = \frac{epM\sin\xi}{(1+e\cos\xi)^2}\dot{\xi}, 
\end{equation}
where $e$ and $p$ are treated as constants when considering only the conservative dynamics. Their time dependence will be restored in the next subsection.

The explicit form of Eqs.~\eqref{subeqn:EOMs-r} and~\eqref{subeqn:EOMs-phi} was first derived in~\cite{Hinderer_2017} and written in terms of the binary’s energy $E$, the azimuthal and radial components of the angular momentum $P_{\phi}, P_r$, and the two potentials $A$ and $D$.
The orbital energy (EOB Hamiltonian) and angular momentum ($P_{\phi}$) are mapped to $e$ and $p$ first by deriving an explicit expression for $P_r$ using Eq.~\eqref{subeqn:EOMs-r}, and then setting it to zero at the turning points $r_{1,2}$.
At the same time, the functions $A$ and $D$ are expanded around the Schwarzschild metric, as explicitly provided in \cite{Buonanno:1998gg, PhysRevD.67.064028, PhysRevD.62.084011, PhysRevD.78.024009}.
By substituting $E, P_{\phi}, P_r, A, D$ in the extended form of Eqs. \eqref{subeqn:EOMs-r}, \eqref{subeqn:EOMs-phi}, and by applying Eq. \eqref{eq:xidot}, we obtain $\dot{\xi}, \dot{\phi}$ as functions of $e,p$:

\begin{subequations}\label{eq:EOMs_cons}
  \begin{align}
    \dot{\xi} &= \frac{(1+e\cos\xi)^2}{Mp^{3/2}}\left[ 1 + \frac{\Xi_{\text{1PN}}}{p} + \frac{\Xi_{\text{2PN}}}{p^2}+\mathcal{O}\left(\frac{1}{p^4}\right)\right] \label{subeqn:EOMs-PN-r}\\
    \dot{\phi} &= \frac{(1+e\cos\xi)^2}{Mp^{3/2}}\left[ 1 + \frac{\Phi_{\text{1PN}}}{p} + \frac{\Phi_{\text{2PN}}}{p^2}+\mathcal{O}\left(\frac{1}{p^4}\right)\right] \label{subeqn:EOMs-PN-phi}
  \end{align}
\end{subequations}
where
\begin{subequations}\label{eq:xi_PN}
  \begin{align}
    \Xi_{\text{1PN}} &= -3(1+e\cos \xi) + \frac{\nu(1-e^2)}{2}\label{subeqn:xi-1PN}\\
    \Xi_{\text{2PN}} &= \frac{1}{8} ( -36+3e^4(\nu-1)\nu + 3\nu(7+\nu) \notag\\
    & + e^2(22 +26\nu-6\nu^2)+ 4e(-2+\nu(11+3e^2))\cos\xi  \notag \\
    &+6e^2(1+2\nu)\cos2\xi)\label{subeqn:xi-2PN}
  \end{align}
\end{subequations}
\begin{subequations}\label{eq:phi_PN}
  \begin{align}
    \Phi_{\text{1PN}} &= -2 e\cos \xi + \frac{\nu(1-e^2)}{2}\label{subeqn:xi-1PN}\\
    \Phi_{\text{2PN}} &= \frac{1}{8} (3(\nu-5)\nu +3e^4(\nu-1) +2e^2(8+\nu-3\nu^2) \notag\\
    &-8e(4+\nu(1-e^2))\cos\xi)\label{subeqn:xi-2PN}.
  \end{align}
\end{subequations}

The geometric interpretation of the two phases is shown in the left plot of \autoref{fig:orbit}.
The orbit is characterized by two frequencies \cite{Hinderer_2017}, $\omega_r$ and $\omega_{\phi}$ associated with the radial and azimuthal phases $\xi, \phi$.
The radial frequency, $\omega_r$, describes the libration between apoapsis and periapsis -- it is the frequency at which the mean anomaly $\psi_r$ evolves -- and is defined as
\begin{equation}\begin{split}
    \omega_r &= \frac{2\pi}{\int_0^{2\pi}d\xi/\dot{\xi}} = \\
    &= \frac{(1-e^2)^{3/2}}{Mp^{3/2}}\left[1 + \frac{\Omega_r^{\text{1PN}}}{p} + \frac{\Omega_r^{\text{2PN}}}{p^2} + \mathcal{O}\left( \frac{1}{p^3}\right)\right]
    \end{split}\end{equation}
where
\begin{subequations}\label{eq:omegar_PN}
  \begin{align}
    \Omega_r^{\text{1PN}} &= \frac{(1-e^2)(-6+\nu)}{2} \label{subeqn:omegar-1PN}\\
    \Omega_r^{\text{2PN}} &= \frac{3(1-e^2)}{8} (8 -20\sqrt{1-e^2} + \nu(-1+8\sqrt{1-e^2} + \nu) \notag \\
    &-e^2(24+ (-5+\nu)\nu))\label{subeqn:omegar-2PN}.
  \end{align}
\end{subequations}

The azimuthal frequency, $\omega_{\phi}$, describes the azimuthal rotational motion, it is the frequency at which the auxiliary variable $\psi_{\phi}$ evolves and it is given as 
\begin{equation}\label{eq:omegaphi}\begin{split}
    \omega_{\phi} &= \frac{\omega_r}{2\pi}\int_0^{2\pi}\frac{\dot{\phi}}{\dot{\xi}}d\xi= \\
    &=\frac{(1-e^2)^{3/2}}{Mp^{3/2}}\left[1 + \frac{\Omega_{\phi}^{\text{1PN}}}{p} + \frac{\Omega_{\phi}^{\text{2PN}}}{p^2} + \mathcal{O}\left( \frac{1}{p^3}\right)\right],
\end{split}\end{equation}
where 
\begin{subequations}\label{eq:phi_PN}
  \begin{align}
    \Omega_{\phi}^{\text{1PN}} &= \frac{\nu+e^2(6-\nu)}{2}\label{subeqn:omegaphi-1PN}\\
    \Omega_{\phi}^{\text{2PN}} &= \frac{\left(e^2 (3-6 \nu )-24 \nu +54\right)}{4}+ \frac{3 \left(1-e^2\right)(\nu -6)}{2}+\notag \\
    &\frac{3 \left(1-e^2\right)}{8}\bigg(8 -20 \sqrt{1-e^2} + \nu  \left(8 \sqrt{1-e^2}+\nu -1\right) \notag \\
    &-e^2 ((\nu -5) \nu +24) \bigg).\label{subeqn:omegaphi-2PN}
  \end{align}
\end{subequations}

The two frequencies coincide at Newtonian order.  For circular binaries ($e=0$),
the two frequencies degenerate and could be combinaed into a single phase. 
For eccentric relativistic binaries, they start to differ from the 1PN order. The difference in periods associated with $\dot{\xi}$ and $\dot{\phi}$ leads to the periapse precession with the phase $\gamma = \phi - \xi$, whose orbital averaged rate is given by $<\dot{\gamma}> = \omega_{\phi} - \omega_r$.
Since we consider mildly relativistic systems (most relevant for PTA), $\gamma$, being the 1PN effect, evolves slower than $\phi$ and $\xi$, and it is convenient to use it as a dynamical variable instead of $\phi$. As we will show later, this allows us to approximate the evolution of $\gamma$ to gain computational efficiency. We also decide to trade the semilatus rectum $p$ for the gauge invariant variable $x = (M \omega_{\phi})^{2/3}$. This change of variable (i) makes easier connection to the circular case \cite{PhysRevD.77.064035}, \cite{Damour:1988mr}, and (ii) is actually measurable from GW signal.
Its explicit form is derived from Eq.~\eqref{eq:omegaphi} and is given by:
\begin{equation}\label{eq:x}
p = \frac{1-e^2}{x}+ p_{1PN} + x\,p_{2PN}
\end{equation}
with
\begin{subequations}\label{eq:phi_PN}
    \begin{align}
    p_{\text{1PN}} &= \frac{1}{3}(\nu - e^2(\nu-6))\label{subeqn:x-1PN}\\
    p_{\text{2PN}} &= \frac{1}{36} \Big[18 -180 \sqrt{1-e^2} +\frac{81-90 \nu }{1+e} +63 \nu \notag\\
    & +72 \sqrt{1-e^2} \nu +4 \nu ^2 +\frac{9 (9-10 \nu)}{1-e} \notag \\
    &-\left(e^2 (\nu  (4 \nu +15)+36)\right) \Big]\label{subeqn:x-2PN}
\end{align}
\end{subequations}

Substituting the semilatus rectum in terms of  $x$ into \eqref{subeqn:EOMs-PN-r} we obtain the evolution of the true anomaly expression in terms of $x$:

\begin{equation}\label{eq:xidot_x}
\begin{split}
\dot{\xi}(x) &= \frac{(1+e \cos\xi)^2x^{3/2}}{M(1-e^2)^{3/2}}\Big[
1 + x\, \Xi_{\text{1PN}}(x) + \\ &x^2\, \Xi_{\text{2PN}}(x)  + \mathcal{O}(x^3)\Big]
\end{split}
\end{equation}

with

\begin{subequations}\label{eq:xi_x_PN}
  \begin{align}
    \Xi(x)_{\text{1PN}} &= -3\frac{(1+ e^2+ e\cos \xi)}{1-e^2} \label{subeqn:xi-x-1PN}\\
    \Xi(x)_{\text{2PN}} &= \frac{1}{4(1-e^2)^2} \left[-48+30\sqrt{1-e^2} -4e^4(\nu-6) +\right.\notag\\& 40\nu 
     -12\nu\sqrt{1-e^2} +2e^2(40-15\sqrt{1-e^2} + \\ &(8+6\sqrt{1-e^2})\nu)   
     \left. -4e(1+e^2(\nu-15)-8\nu)\cos\xi \right.\notag \\ & \left.+3e^2(1+2\nu)\cos 2\xi\right].\label{subeqn:xi-x-2PN}
  \end{align}
\end{subequations}

Similarly, the orbit-averaged expression for the precession phase in terms of $x$ can be written as 
\begin{equation}\begin{split}\label{eq:gammadotx}
   <\dot{\gamma}> &= \omega_r\left[\int_{0}^{2\pi}\frac{\dot{\phi}}{2\pi \dot{\xi}} d\xi -1 \right] = \frac{3x^{5/2}}{(1-e^2)M}\\
   & + \frac{(18-21e^2-28\nu-2e^2\nu)x^{7/2}}{4M(e^2-1)^2},
\end{split}\end{equation}
which depends explicitly on the total mass $M$ of the binary. 

\subsection{\label{sec:dissipative}Dissipative Dynamics}

A binary system radiates orbital energy and angular momentum through the emission of GWs. This dissipation causes the binary orbit to shrink and gradually circularize until the system merges.
We derive the dissipative terms in Eqs.~\eqref{subeqn:EOMs-Pr} and \eqref{subeqn:EOMs-Pphi} following \cite{Hinderer_2017}, where the orbit-averaged fluxes $\langle \mathcal{F} \rangle$ and $\langle \mathcal{G} \rangle$ are given as

\begin{widetext}
\begin{align}
\biggl\langle \frac{dE}{dt} \biggr\rangle
&= - \langle \mathcal{F} \rangle \notag\\
&= - \Biggl[
  \frac{(37 e^4+292 e^2+96)\nu^2 x^5}{15 (1-e^2)^{7/2}}
  - \frac{x^6 \nu^2}{2520 (1-e^2)^{9/2}}
    \Bigl(
      e^6 (5180\nu+36333)
      + 42 e^4 (3520\nu+9253) \notag\\
&\qquad\qquad\qquad
      + 8 e^2 (34160\nu+47703)
      + 48 (980\nu+1247)
    \Bigr)
  + \mathcal{O}(x^{13/2})
\Biggr]
\label{eq:average_energy_flux}
\\[1.2em]
\biggl\langle \frac{dP_{\phi}}{dt} \biggr\rangle
&= - \langle \mathcal{G} \rangle \notag\\
&= - \Biggl[
  \frac{4 (7 e^2 + 8)\, \nu^2 M x^{7/2}}
       {5 (e^2 - 1)^2}
  + \frac{\nu^2 M x^{9/2}}
         {420 (e^2 - 1)^3}
    \Bigl(
      e^4 (2996\nu + 5713)
      + 8 e^2 (2758\nu + 2777)
      + 7840\nu + 9976
    \Bigr) \notag\\
&\quad\quad\quad\quad
  + \frac{\pi}{5}
    (2415 e^4 + 836 e^2 + 128)\,
    \nu^2 M x^5
  + \mathcal{O}\!\left(x^{11/2}\right)
\Biggr]
\label{eq:average_angular_momentum_flux}
\end{align}
\end{widetext}

We then incorporate these dissipative effects into the evolution of our dynamical variables, $e$ and $x$, which were held constant in previous subsevtion. These quantities are directly related to the energy and angular-momentum fluxes as follows

\begin{widetext}
\begin{eqnarray}
\label{eq:de,dx/dt}
    \frac{de}{dt} &=& \frac{((dP_{\phi}/dt)(\partial E/\partial x) - (dE/dt)(\partial P_{\phi}/\partial x))(\partial x/\partial p)}{{((\partial P_{\phi}/\partial e)(\partial E/\partial x) - (\partial P_{\phi}/\partial x)(\partial E/\partial e))(\partial x/\partial p)}}
    \nonumber\\
    \frac{dx}{dt} &=& \frac{\partial x}{\partial p}\left(\frac{(dE/dt)(\partial P_{\phi}/\partial e) - (dP_{\phi}/dt)/(\partial E/\partial e) }{{((\partial P_{\phi}/\partial e)(\partial E/\partial x) - (\partial P_{\phi}/\partial x)(\partial E/\partial e))(\partial x/\partial p)}}\right)\nonumber\\
\end{eqnarray}
\end{widetext}

After explicit substituting we obtain
\begin{subequations}\label{eq:de,dx/dt_final}
  \begin{align}
    \frac{de}{dt} &= x^4\, E_{\text{2.5PN}} + x^5\, E_{\text{3.5PN}}  + \mathcal{O}(x^{11/2}), \\
    \frac{dx}{dt} &= x^5\, X_{\text{2.5PN}}  + x^6\, X_{\text{3.5PN}}  + \mathcal{O}(x^{13/2}),
  \end{align}
\end{subequations}
where
\begin{subequations}\label{eq:EPN}
  \begin{align}
    E_{\text{2.5PN}} &= -\frac{e\nu \left(121 e^2+304\right)}{15 \left(\left(1-e^2\right)^{5/2} M\right)}, \label{subeqn:E-2.5PN}\\
    E_{\text{3.5PN}} &= \frac{e \nu}{2520 \left(1-e^2\right)^{7/2} M} \bigg(e^4 (19768 \nu +94887)+ \notag \\
    & + 12 e^2 (21427 \nu +38698)+8 (24556 \nu +20547)\bigg)\label{subeqn:E-3.5PN},
  \end{align}
\end{subequations}
\begin{subequations}\label{eq:XPN}
  \begin{align}
    X_{\text{2.5PN}} &= \frac{2\nu \left(37 e^4+292 e^2+96\right)}{15 \left(1-e^2\right)^{7/2} M},\label{subeqn:X-2.5PN}\\
    X_{\text{3.5PN}} &= -\frac{\nu}{420 \left(\left(1-e^2\right)^{9/2} M\right)}  \bigg(e^6 (2072 \nu +6931)+ \notag \\
    & +14 e^4 (3690 \nu +7079) +8 e^2 (11158 \nu +15411) +\notag \\
    &+16 (924 \nu +743)\bigg). \label{subeqn:X-3.5PN}
  \end{align}
\end{subequations}

\section{\label{sec:waveform}Gravitational Waveform}
The GW strain  in the source reference frame can be written as a decomposition in spherical harmonic modes $h^{\ell m}$ \cite{RevModPhys.52.299}:
\begin{equation}
    h =  h_+ - ih_{\times} = \sum_{\ell = 0}^{\infty} \sum_{m=-\ell}^{+\ell} {}_{-2}Y^{\ell m}(\Theta, \Phi)h^{\ell m},
\end{equation}
and explicit form of the spin-weighted spherical harmonics of spin weight $s=-2$, ${}_{-2}Y^{\ell m}(\Theta, \Phi)$ can be found in \cite{PhysRevD.91.084040}.
We introduce a radiation frame by associating the polar angle with the inclination of the orbital angular momentum to the line of sight $\Theta = \iota$. We also choose the $\rm{\bf{x}}$-axis of the source frame such that the observer lies in the $\rm{\bf{z}}$–$\rm{\bf{x}}$ plane implying $\Phi=0$, which leads to
\begin{equation}\label{eq:Y2-20}
{}_{-2}Y^{2\pm2} = \sqrt{\frac{5}{64\pi}}(1 \pm \cos\iota)^2,\hspace{0.2cm} {}_{-2}Y^{20} = \sqrt{\frac{15}{32\pi}}\sin^2\iota. 
\end{equation}

For the inspiral phase, the sum can be truncated at the dominant modes $(\ell = 2, \hspace{0.1cm} m = 0,\pm2)$. The $m=\pm 1$ mode is suppressed by a PN factor $v/c$ where $v$ is an orbital velocity. Moreover, this mode is also proportional to  the mass difference between two companions, which is small since we consider equal (or comparable) mass binaries. Note that the dynamical part of the $m=0$ mode is proportional to the eccentricity and is therefore usually neglected in the circular case.

We consider only the instantaneous part of the waveform written as  
\begin{equation}
    h^{2, m}_{\text{inst}}  = \frac{4M\nu}{R} \sqrt{\frac{\pi}{5}} e^{-i m\phi} \hat{H}^{2, m}_{\text{inst}}
\end{equation}
and expressed in terms of the dynamical variables $(r, \dot{r}, \phi, \dot{\phi})$, truncated at Newtonian order \cite{PhysRevD.91.084040}:
\begin{subequations}\label{eq:H220-2}
  \begin{align}
    &\hat{H}^{20} = -\sqrt{\frac{2}{3}}\left( \frac{M}{r} -r^2\dot{\phi}^2 -\dot{r}^2\right) \label{subeqn:H20}\\
    &\hat{H}^{22} = \frac{M}{r} + r^2\dot{\phi}^2 + 2i r\dot{r}\dot{\phi} -\dot{r}^2\label{subeqn:H22}\\
    &\hat{H}^{2-2} = \left(H^{22}\right)^{*}\label{subeqn:H2-2}
  \end{align}
\end{subequations}

We need to solve the equations of motion and substitute solution in these expressions, then we obtain explicitly: 
\begin{widetext}
\begin{subequations}\label{eq:h+,x}
  \begin{align}
    h_{+} &= \frac{M\nu x}{R(1-e^2)}\bigg\{(1+\cos^2\iota)\left[(2+3e\cos\xi +e^2\cos2\xi)\cos 2\phi + (2e\sin\xi +e^2\sin 2\xi)\sin 2\phi\right] +\sin^2\iota \left[e(e+\cos\xi)\right]\bigg\}, \\
    h_{\times} &= \frac{M\nu x}{R(1-e^2)}\bigg\{2\cos\iota\left[(2+3e\cos\xi +e^2\cos2\xi)\sin 2\phi - (2e\sin\xi +e^2\sin 2\xi)\cos 2\phi\right]\bigg\},
  \end{align}
\end{subequations}
and replacing the azimuthal phase $\phi$, with the precession phase $\gamma$ gives us
\begin{subequations}\label{eq:h+,x_new}
  \begin{align}
    h_{+} = \frac{M\nu x}{R(1-e^2)}\bigg\{&(1+\cos^2\iota)\left[\left(e^2 +2\cos 2\xi +\frac{5}{2}e\cos\xi +\frac{e}{2}\cos 3\xi\right)\cos 2\gamma - \left(\frac{5}{2}e\sin\xi + 2\sin 2\xi +\frac{e}{2}\sin 3\xi\right)\sin 2\gamma\right] \notag \\
    &+\sin^2\iota \left[e(e+\cos\xi)\right]\bigg\}, \\
    h_{\times} = \frac{M\nu x}{R(1-e^2)}\bigg\{&2\cos\iota\left[\left(e^2 +2\cos 2\xi +\frac{5}{2}e\cos\xi +\frac{e}{2}\cos 3\xi\right)\sin 2\gamma + \left(\frac{5}{2}e\sin\xi + 2\sin 2\xi +\frac{e}{2}\sin 3\xi\right)\cos 2\gamma\right]\bigg\}.
  \end{align}
\end{subequations}
\end{widetext}
There are three time-scales entering the phase of GW signal: (i) orbital time scale in $\xi(t)$ (ii) precession timescale $\gamma(t)$ which is significantly longer than orbital for binaries in PTA band (iii) radiation-reaction time-scale $e(t), x(t)$ which is also much longer than orbital period. Note, that the waveform written in this form can be re-written in harmonic form with the phase  $(\xi, k\xi \pm 2\gamma)$, where $k=\pm(0,1,2,3)$.

\section{\label{sec:residuals}Residuals}
The passage of a GW affects the geodesic trajectory of photons propagating from the pulsar to Earth. It is convenient to use a transverse-traceless coordinate frame to describe the geodesic deviation, where the interaction of e/m radiation with GW can be seen as a Doppler modulation of radio pulses observed on Earth. In other words, we observe shifts in the Time of Arrival (ToA) of the pulsar signal relative to the expected time intervals defined by the rotational period of the pulsar $T_{\alpha}$. Mathematically, this can be expressed as \cite{Sesana_Vecchio2010, PhysRevD.85.044034}  

\begin{equation}\begin{split}
    z_{\alpha}(t) &\equiv \left( \frac{\nu_0 - \nu(t)}{\nu_0}\right)_{\alpha} \\ &=F^{+,\times}_{\alpha}\left[ h_{+,\times}(t, \mathbf{x} = 0) - h_{+,\times}(t - \tau_{\alpha}, \mathbf{x}_{\alpha})\right],
\end{split}\end{equation}
where we assume the sum over GW polarisations $(+, \times)$; $t$ and $\mathbf{x} = 0$ correspond to the observer’s time and position, $\mathbf{x}_{\alpha}$ is the position of the pulsar, and $\tau_{\alpha}$ is the light travel time from the pulsar to Earth as given by~\cite{Sesana_Vecchio2010}:
\begin{eqnarray}\label{eq:tau}
\tau_{\alpha} &=& L_{\alpha}\left(1+\hat{\Omega}_{\text{GW}}\cdot\hat{p}_{\alpha}\right) \nonumber \\
&\simeq& 1.1 \times 10^{11}\frac{L_p}{1\text{kpc}}\left(1+\hat{\Omega}_{\text{GW}}\cdot\hat{p}_{\alpha}\right)\text{s} \nonumber \\
&\simeq& 3.5 \times 10^3\frac{L_p}{1 \text{kpc}}\left(1+\hat{\Omega}_{\text{GW}}\cdot\hat{p}_{\alpha}\right)\text{yr},
\end{eqnarray}
where $\alpha$ is the pulsar index, $\hat{\Omega}_{\text{GW}}$ is a unit vector in the direction of GW propagation and $\hat{p}_{\alpha}$ is a direction to the pulsar. 
The antenna pattern functions $F_{\alpha}^{+, \times}$ for the $\alpha$-th pulsar are defined as \cite{Taylor_2016, Ellis_2012}:

\begin{equation}\begin{split}
F^{+}_{\alpha} &= \frac{1}{2} 
\frac{(\hat{m}\cdot \hat{p}_{\alpha})^2 - (\hat{n}\cdot \hat{p}_{\alpha})^2}
     { 1 + \hat{{\Omega}}\cdot \hat{p}_{\alpha} } \nonumber \\
F^{\times} &= 
\frac{ (\hat{m}\cdot \hat{p}_{\alpha})(\hat{n}\cdot \hat{p}_{\alpha}) }
     { 1 + \hat{\Omega}\cdot \hat{p}_{\alpha} }  
\end{split}\end{equation}

where $\hat{p}_{\alpha}$ is a unit vector defining the position of a pulsar in the sky and $\hat{\Omega}$ is the unit vector defining the direction of propagation of the GW:
\begin{equation}\begin{split}
\hat{m} &= (-\sin\phi_{\mathrm{GW}}, \, \cos\phi_{\mathrm{GW}}, \, 0), \\
\hat{n} &= (-\cos\theta_{\mathrm{GW}} \cos\phi_{\mathrm{GW}}, \, 
              -\cos\theta_{\mathrm{GW}} \sin\phi_{\mathrm{GW}}, \, 
               \sin\theta_{\mathrm{GW}}), \\
\hat{\Omega}_{\text{GW}} &= (-\sin\theta_{\mathrm{GW}} \cos\phi_{\mathrm{GW}}, \,
                              -\sin\theta_{\mathrm{GW}} \sin\phi_{\mathrm{GW}}, \,
                              -\cos\theta_{\mathrm{GW}})
\end{split}\end{equation}

In practice, PTA observations work with the ``phase of arriving pulses'' -- that is, the difference between the expected and measured number of pulses during the observation time divided by a pulsar spin frequency. The residuals are, therefore, given as 
\begin{equation}
    r(t) = \int_0^tdt' z(t', \hat{\Omega}).
\end{equation}
or, explicitly for each pulsar in the array,
\begin{equation}\begin{split}
    r_{\alpha}(t) &= r^{(E)}_{\alpha}(t) - r^{(p)}_{\alpha} (t) = \\ &[F^{+}_{\alpha}\cos(2\psi) + F^{\times}_{\alpha}\sin(2\psi)] (r_{+}(t) -  r_{+}(t-\tau_{\alpha})) - \\&[F^{+}_{\alpha}\sin(2\psi) - F^{\times}_{\alpha}\cos(2\psi)](r_{\times}(t) -  r_{\times}(t-\tau_{\alpha}))
\end{split}\end{equation}

where $r_{+,\times}(t) = \int_0^tdt' h_{+,\times}(t')$.
The response in residuals to GW consists of two terms  which we mentioned above and can explicitly introduce now. The first, \emph{Earth} term, $r^{(E)}$, is evaluated at the time of reception of the radio pulses, and it is coherent across all the pulsars. The second, \emph{pulsar} term, $r^{(p)}$, is incoherent, as it depends on the distance to the pulsar and appears at different frequencies in each pulsar's data.  

To find the residuals, we integrate ~\eqref{eq:h+,x_new} neglecting time evolution of
$e, x, \gamma$ and restoring their time dependence in the final expression. This is similar to how we obtained $\xi(t)$ in \eqref{eq:xi_of_t}, and here we can use the same justification. 
The residuals corresponding to two GW polarisations are given explicitly as 
\begin{widetext}
\begin{subequations}\label{eq:r+,x}
  \begin{align}
    r_{+} &= \frac{\nu M^2 (1 - e^2)^{3/2}}{R \sqrt{x}(1 - e^2 - 3 x)}\biggl\{(1+\cos^2\iota)\biggl[\left( \frac{(e + 2\cos \xi) \sin\xi}{1+e \cos\xi}\right)\cos(2\gamma) + \left(\frac{\cos2\xi+e\cos\xi}{1+e\cos\xi}\right)\sin(2\gamma)\biggr] + \sin^2\iota \left( \frac{e \sin \xi}{1+e \cos\xi}\right)\biggr\}, \\
    r_{\times} &=\frac{\nu M^2 (1 - e^2)^{3/2}}{R \sqrt{x}(1 - e^2 - 3 x)} \biggl\{2\cos\iota\biggl[\left( \frac{(e + 2\cos \xi) \sin\xi}{1+e \cos\xi}\right)\sin(2\gamma)-\left(\frac{\cos2\xi+e\cos\xi}{1+e\cos\xi}\right)\cos(2\gamma)\biggr] \biggr\}.
  \end{align}
\end{subequations}
\end{widetext}

\section{\label{sec:results}Results}
\subsection{\label{sec:solution_ODEs}Binary evolution}

Building the search waveform for PTA consists of three steps.
First, we need to integrate the binary dynamics; second, we compute the GW strain; and finally, we include the PTA response -- we compute 
 the timing residuals produced by the passing GW signal.

In this subsection, we focus on the first step -- solving the system of ODEs given in Eqs.~\eqref{eq:xi_x_PN}, ~\eqref{eq:gammadotx} and~\eqref{eq:de,dx/dt_final}. These equations can be solved numerically, and we need to consider two characteristic timescales:
(i) the radio observation time of each pulsar (typically between 5 and 20 years) and
(ii) the light-travel time between the pulsar and Earth (typically about a thousand years).

The first time scale describes the GW signal during the PTA observation period, while the second is required to compute the \emph{pulsar} term \cite{10.1111/j.1365-2966.2011.18622.x}.
The start of the pulsar term is delayed by $\tau_{\alpha}$ relative to the Earth term and therefore requires the GW signal at the time of radio emission. 

Numerical integration of the binary dynamics over $\tau_{\alpha}$ to compute the pulsar term is unavoidable and currently represents the most computationally expensive part of waveform generation.

\subsubsection{Approximate solution over the observation time}\label{sec:dynamics_approx}

Integration of the dynamics over the observation time, however, can be approximated -- this is the first goal of this paper.
We evaluate the accuracy of this approximation by comparing it to the full numerical solution of the ODEs.

Since SMBHBs in the PTA band evolve only slowly over the observation timescale ($T_{\rm obs}$), we first truncate the equations for $e$ and $x$ (Eq. ~\eqref{eq:de,dx/dt_final}) at leading order and then use a Taylor expansion, retaining only the linear evolution in time:
 
  \begin{subequations}\label{eq:EPN1, eq:EPN2}
  \begin{align}
    e(t) &= e(t_0) + E_{\text{2.5PN}}(t_0) t + \mathcal{O}(t^2)\label{subeqn:e_approx}\\
    x(t) &= x(t_0) + X_{\text{2.5PN}}(t_0) t + \mathcal{O}(t^2)\label{subeqn:x_approx}
   \end{align}
\end{subequations}

As noted previously, phase $\gamma = \phi - \xi$ characterises the relativistic precession of the periapsis, which arises at the first post-Newtonian (1PN) order. We can neglect all oscillations in its temporal evolution and use an orbit-averaged, linear-in-time approximation truncated at 1PN order:
\begin{equation}
\label{eq:gammadot}
    \gamma(t) = \gamma(t_0) + <\dot{\gamma}>(t_0)(t) + \mathcal{O}(t^2),
\end{equation}
where $ <\dot{\gamma}> = \langle\dot{\phi} - \dot{\xi}\rangle$  and explicitely given as
\begin{eqnarray}
<\dot{\gamma}> =  \omega_{\phi}^{\text{1PN}} - \omega_r ^{\text{1PN}}
\end{eqnarray}

\begin{figure}[tbh]
    \centering
    \includegraphics[width=0.5\textwidth]{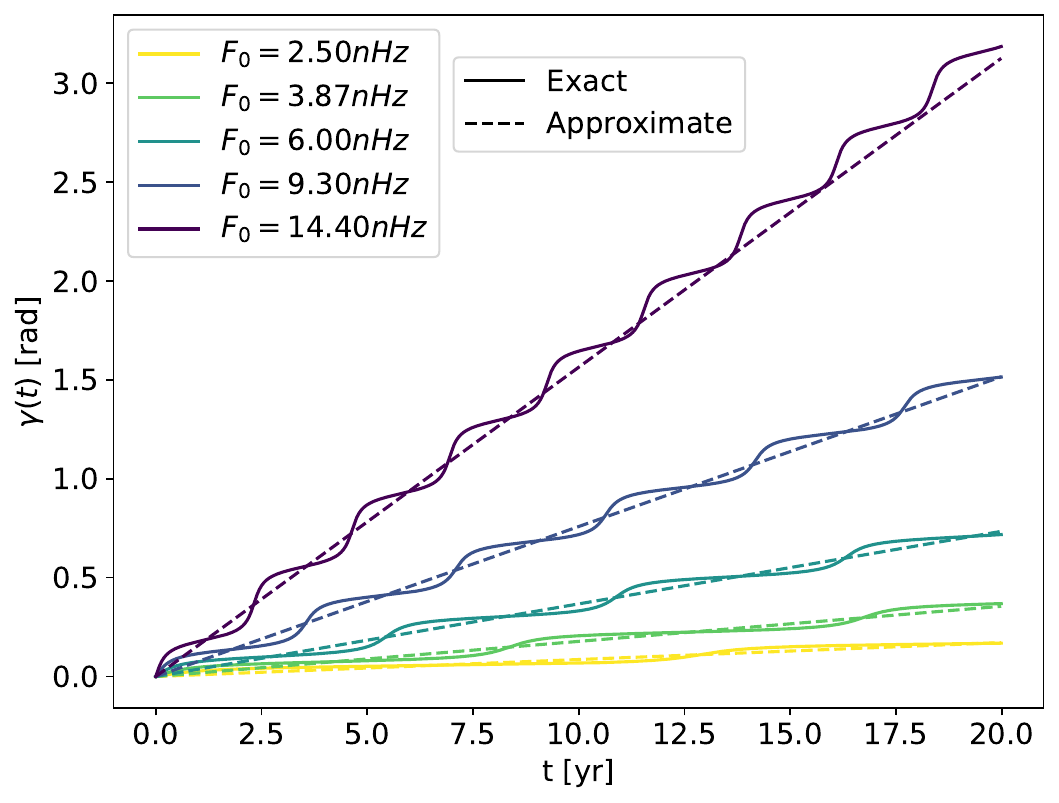}
    \caption{Evolution of $\gamma(t)$ as a function of time computed over $20 yrs$. Different colors correspond to different frequencies, while initial eccentricity is fixed to $e_0 = 0.5$ and $\mathcal{M}_c = 10^{9.2}M_{\odot}$. Approximate solution $<\gamma(t)>$, is dashed and compared with the ``exact'' time integral of $\dot{\gamma} = \dot{\phi} - \dot{\xi}$.}
    \label{fig:gamma_t}
\end{figure}

We compare the approximate and exact evolution of the precession phase $\gamma(t)$ in  \autoref{fig:gamma_t}.

The orbital motion (evolution on the  orbital time scale) is encoded in $\xi(t)$. We solve instead  
$$
t(\xi) - t(\xi_0) = \int_{\xi_0}^{\xi}{\frac{1}{\dot{\xi}(\xi')} d\xi'}
$$
assuming that $x, e$ are constant and restoring the time evolution of $x(t), e(t)$ in the final expression:  

\begin{widetext}
\begin{align}
\label{eq:xi_of_t}
t(\xi) &= t(\xi_0)
+ \frac{2 M\, p^{5/2}}{(2p + \nu - \nu e^2)^2}
\Bigg[
\frac{2 \big(6 + 2p + \nu - e^2 (6 + \nu)\big)}
{(1 - e^2)^{3/2}}\,
\tan^{-1}\!\!\left(
\sqrt{\frac{1-e}{1+e}}\,
\tan\!\frac{\xi}{2}
\right)
\nonumber\\[8pt]
&\quad
-\frac{72\,
\tan^{-1}\!\!\left(
\sqrt{\frac{6 - 2p - \nu - e(6 - e\nu)}
{6 - 2p - \nu + e(6 + e\nu)}}\,
\tan\!\frac{\xi}{2}
\right)}
{\sqrt{\big(-6 + 2p + \nu + e(6 - e\nu)\big)
\big(-6 + 2p + \nu - e(6 + e\nu)\big)}}-\frac{e\,(2p + \nu - e^2 \nu)\, \sin\xi}
{(1 - e^2)(1 + e\cos\xi)}
\Bigg],
\end{align}
\end{widetext}
The resulting function is approximated by a spline constructed on a uniform, finely sampled grid in $\xi$, which we then invert to recover $\xi(t)$.
The approximate set of equations is then summarized in Eqs.~\eqref{eq:xi_of_t}, \eqref{eq:gammadot}, and \eqref{eq:EPN1, eq:EPN2}.

We are now ready to assess the accuracy of our approximation.
For validation purposes, we fix the binary’s chirp mass to
$\mathcal{M}_c = 10^{9.2}M_{\odot}$. Note that the orbital parameter $x = (M \omega_{\phi})^{2/3}$ depends both on the total mass $M$ and the azimuthal frequency $\omega_{\phi}$.
We assume an equal-mass binary (symmetric mass ratio $\nu = 0.25$), then variations in $x$ can be interpreted as arising from changes in either the total mass or the orbital frequency.

We also emphasize that, at the leading post-Newtonian (PN) order, only the evolution of the precession phase $\gamma$ depends explicitly on the total mass; all other equations can be rewritten solely in terms of the chirp mass.
The chirp mass is, in fact, the best-measured parameter when only the inspiral portion of the gravitational-wave (GW) signal is observed.

We have already demonstrated agreement between the approximate (linear) evolution of the precession phase and its ``exact'' numerical counterpart in \autoref{fig:gamma_t}.  Our approximation captures well the secular evolution, but, being averaged over the orbit, does not reproduce the oscillatory behaviour. Such deviations may become relevant only for very high–SNR GW detections, which are beyond the reach of current PTA sensitivities.

\begin{figure}[h] 
    \centering
    \includegraphics[width=0.45\textwidth]{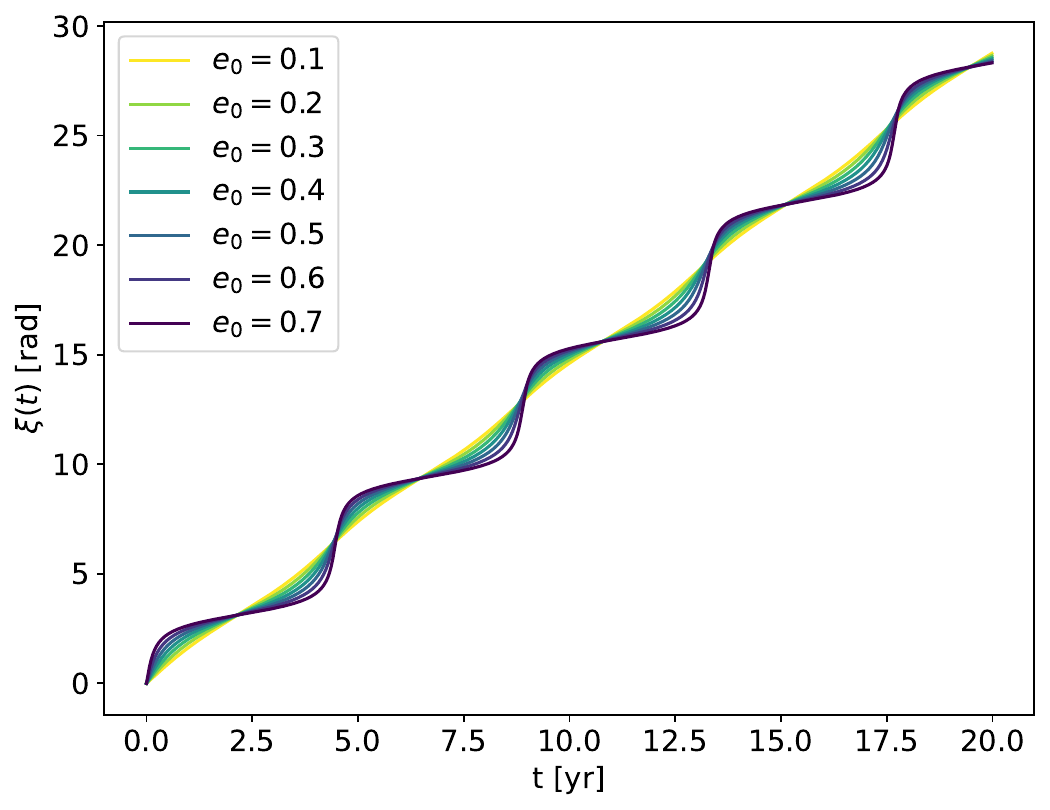}
    \caption{Evolution of $\xi(t)$ as a function of time computed over $20 yrs$. Different colors correspond to different initial eccentricities, while initial azimuthal orbital frequency is fixed to $F_{\text{orb}} = 7.5 \text{nHz}$ and $\mathcal{M}_c = 10^{9.2}M_{\odot}$.}
    \label{fig:xi_t}
\end{figure}

In \autoref{fig:xi_t}, we show the behaviour of $\xi(t)$ over a 20-year observation period, $T_{\rm obs} = 20~\text{yrs}$, for a binary starting at $F_{\rm orb} = \omega_{\phi}/(2\pi) = 7.5 \text{nHz}$ and for different initial eccentricities.
In deriving $t(\xi)$ from equation \eqref{eq:xi_of_t}, the integration is performed under the assumption that $e$ and $x$, remain constant and their temporal evolution is replaced afterwards.
As eccentricity increases, $\xi(t)$ approaches a step-like behaviour, reflecting the increasingly asymmetric orbital motion near periapsis. For $e_0 > 0.75$ and $F_{\text{orb}}> 20 nHz $, this behaviour introduces numerical instabilities when inverting the spline to recover $\xi(t)$, which require careful treatment. 

\begin{figure}[h] 
    \centering
    \includegraphics[width=0.45\textwidth]{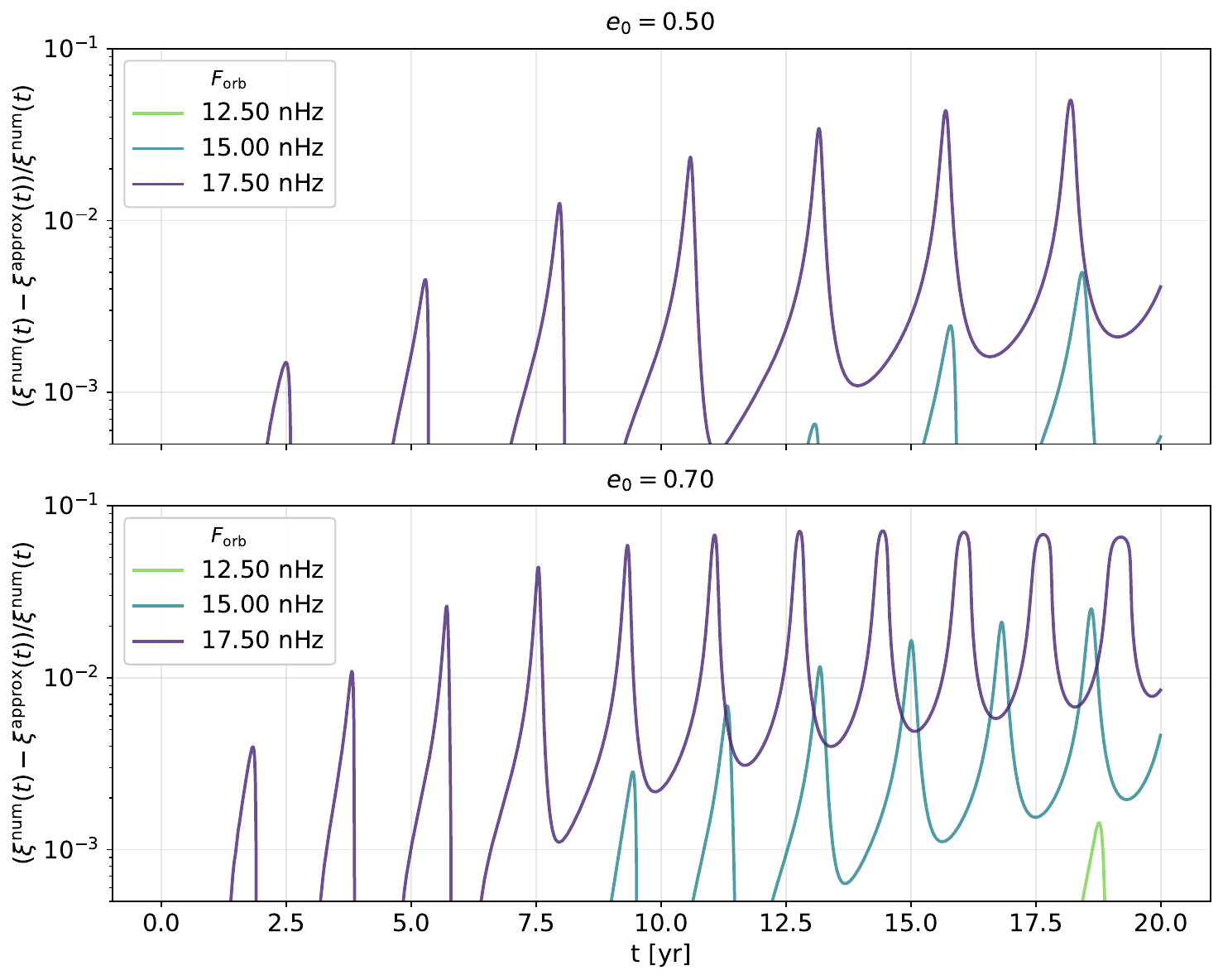}
    \caption{Discrepancy between numerical and approximate solution of $\xi(t)$ as a function of time computed over $20 yrs$. Different colors correspond to different orbital frequencies and upper panel is for $e_0 = 0.5$ and lower panel for $e_0 = 0.7$.}
    \label{fig:diff_xi_t}
\end{figure}

Comparing the numerical and approximate solutions for $\xi(t)$, we obtain excellent agreement for orbits with $e_0 < 0.5$ and $F_{\text{orb}} < 7,\mathrm{nHz}$, with discrepancies below numerical precision.
As eccentricity and frequency increase, the difference between the two solutions becomes progressively more pronounced over time.
In \autoref{fig:diff_xi_t}, we show the relative difference between the numerical and approximate solutions, $(\xi^{\text{num}}(t) - \xi^{\text{approx}}(t))/\xi^{\text{num}}$.
For highly eccentric binaries at low frequency, the error accumulates over time and becomes significant by the end of the observation (green line in the bottom panel of \autoref{fig:diff_xi_t}).
In contrast, at higher frequencies, the discrepancy is visible throughout the observation period for initial eccentricities $\gtrsim 0.5$.
The periodic peaks correspond to the step-like behaviour of $\xi(t)$ near periapsis.
The examples shown represent the boundary of the parameter space where the approximate solution remains valid.

\begin{figure}[h]
    \includegraphics[width=0.5\textwidth]{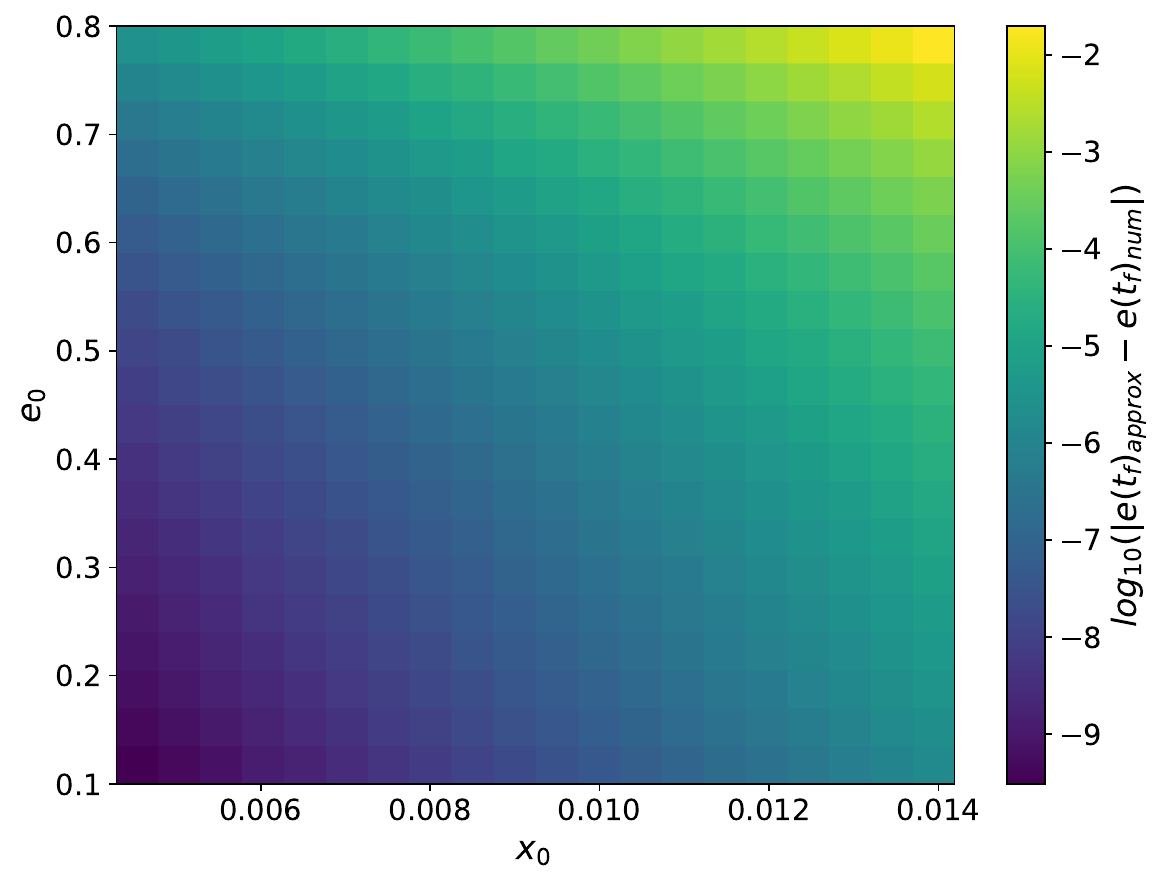}
    \caption{Difference between numerical and approximate evolution of $e$ for a SMBHB with $\mathcal{M}_c = 10^{9.2}$ on a timespan of $20$ yrs, computed at different values of $e_0$ and $x_0$.}
    \label{fig:err_e}
\end{figure}

\begin{figure}[h]
    \includegraphics[width=0.5\textwidth]{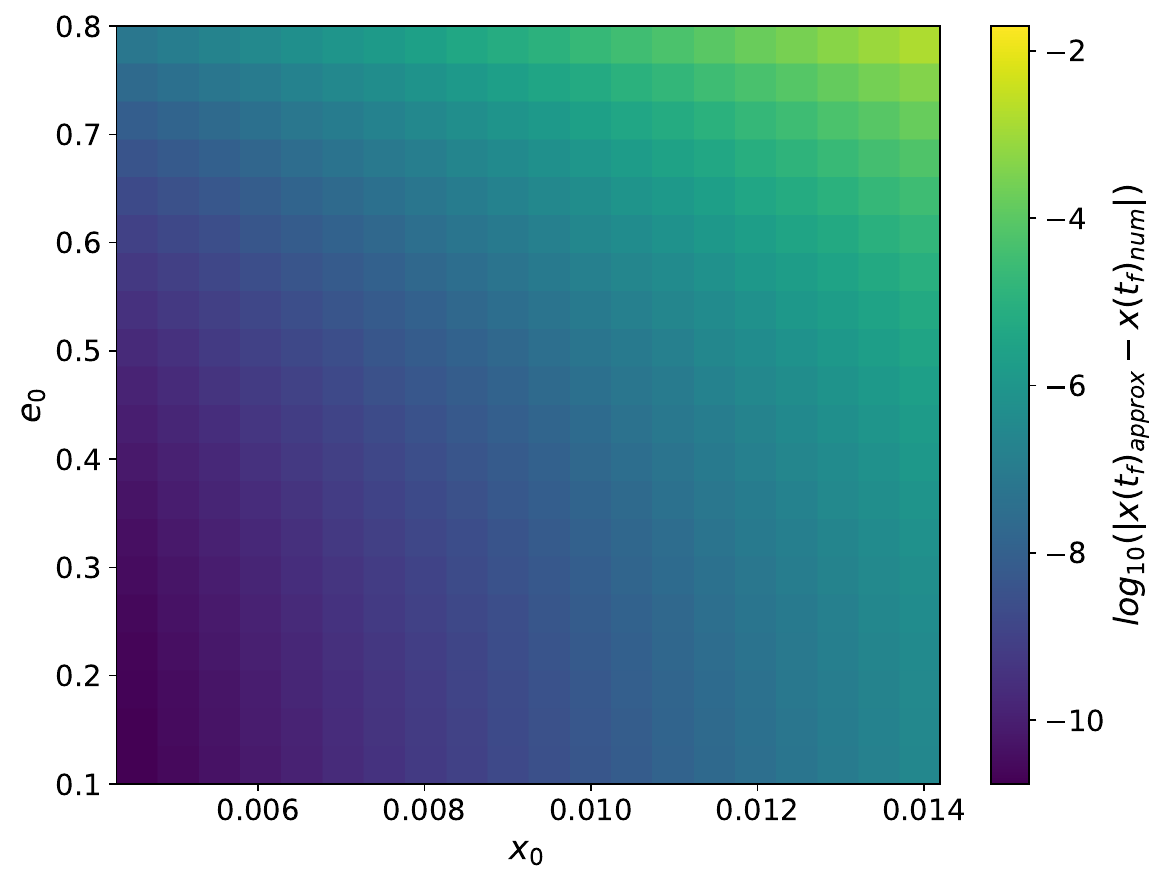}
    \caption{Difference between numerical and approximate evolution $x$ for a SMBHB with $\mathcal{M}_c = 10^{9.2}$ on a timespan of $20$ yrs, computed at different values of $e_0$ and $x_0$.}
    \label{fig:err_x}
\end{figure}

In Figures \ref{fig:err_e}, \ref{fig:err_x}, we compare the approximate and fully numerical evolutions of $e(t)$ and $x(t)$ over $T_{\rm obs} = 20~\text{yrs}$ for a range of initial orbital parameters $(e_0, x_0)$. The evolution is compared for different initial
orbital elements $e_0, x_0$. The colour map represents the logarithm of the absolute difference evaluated at the end of the integration, where the deviations are expected to reach their maximum.
As anticipated, the strongest discrepancies appear for binaries with high orbital frequencies and large eccentricities -- those that are more relativistic and experience faster orbital decay and circularization. Nevertheless, even in these cases, the approximation introduced in this subsection maintains sub-percent accuracy, validating its suitability for PTA waveform modelling. Going to eccentricity $e_0>0.8$ and higher values of $x_0\gtrsim 0.014$, which imply larger masses and higher orbital frequency, require numerical integration of the ODEs to maintain the necessary accuracy, 
this is the subject of next subsection.

\subsubsection{Numerical integration}\label{sec:dynamics_num}

\begin{figure}
\centering
\includegraphics[width=0.5\textwidth]{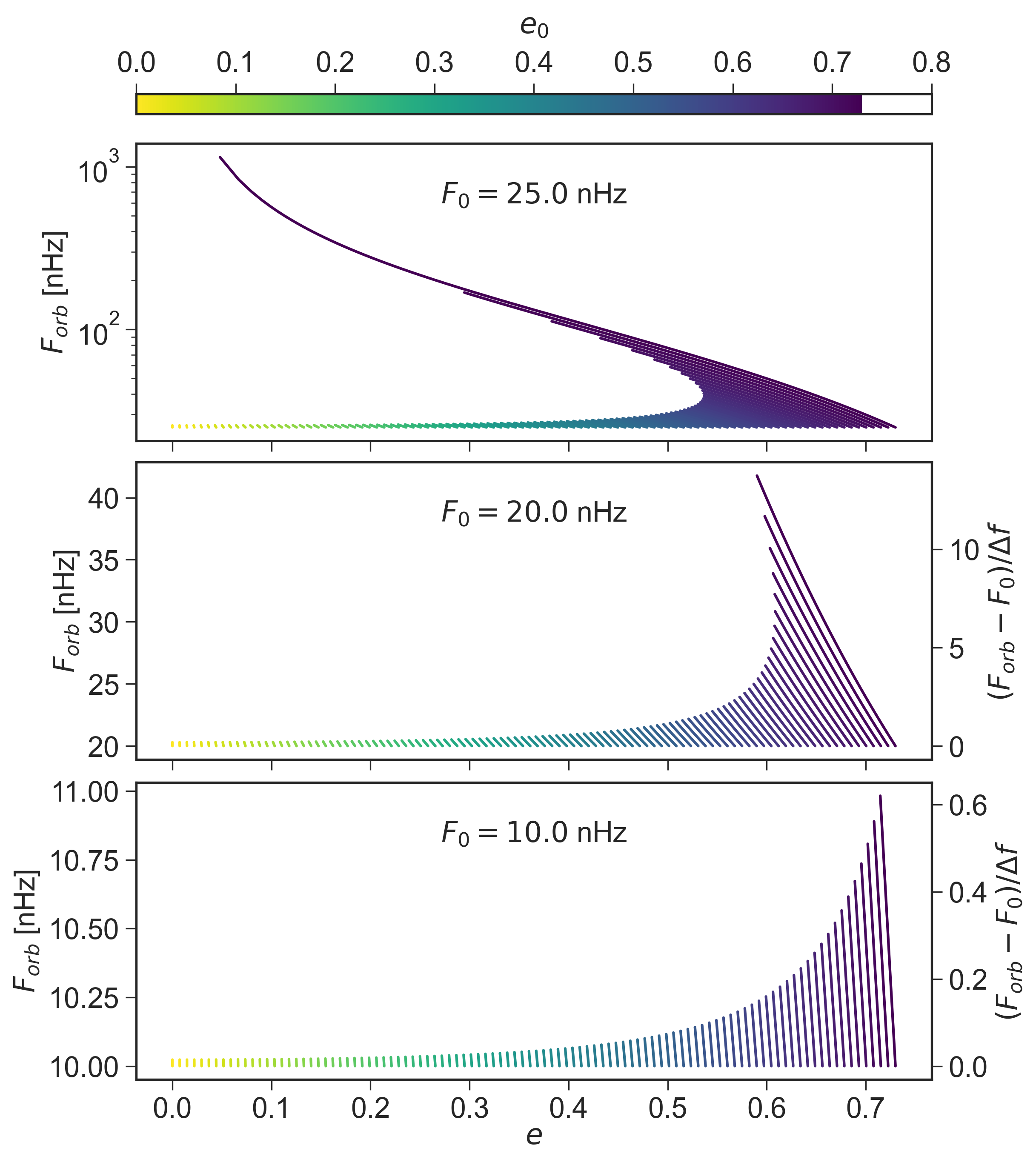}
\caption{\label{fig:evolution_eF} Evolution of eccentricity and orbital frequency as a function of $e_0$. The colorbar represents different values of $e_0$, for each we compute the evolution over $20 yrs$ in $F_{orb}$ (y-axis) and $e$ (x-axis).}
\end{figure}

High-mass, high-frequency binaries are expected to be rare in the PTA band, as they evolve relatively quickly \cite{leclere2025multimessengerviewpulsartiming}; however, they also represent the most promising candidates for individually detectable sources.
In \autoref{fig:evolution_eF}, we show the orbital evolution of our fiducial equal-mass binary with $\mathcal{M}_c = 10^{9.2}M_{\odot}$ over 20 years of observation. Each panel corresponds to a different starting orbital frequency,
$F_{\rm{orb}} = x^{3/2}/(2\pi M) = {10, 20, 25}~\text{nHz}$, and presents the binary’s evolution in the $(F_{\rm{orb}}, e)$ plane as a function of its initial eccentricity.

The bottom panel corresponds to binaries for which the approximation described in the previous subsection remains valid.
We observe an approximately linear behaviour; however, even in this case, high-eccentricity orbits exhibit significant evolution (the right-hand scale shows the evolution in units of Fourier bins).
The middle panel illustrates a regime where the linear approximation begins to break down at high eccentricities, and the effects of orbital circularization become apparent.
Finally, the top panel shows a binary that “chirps” across and eventually leaves the PTA band.
Most binaries shown in this panel require full numerical integration of the ODEs.
We emphasize that the behaviour of eccentric binaries in the PTA band differs substantially from that of circular SMBHBs.

As already mentioned, obtaining the orbital parameters at the pulsar time requires evolving the orbit backward in time over $\sim 10^3$ years.
In addition to the need for numerically solving the ODEs, we also must include the 2PN terms in the dynamics.
We do not claim to have sufficient accuracy to determine the pulsar phase $\xi(-\tau_{\alpha})$, especially given the poor knowledge of pulsar distances.
Therefore, we evolve only $e$, $x$, and $\gamma$ backward in time, treating the pulsar phase as an independent parameter.

\begin{figure}
\centering

\includegraphics[width=0.5\textwidth]{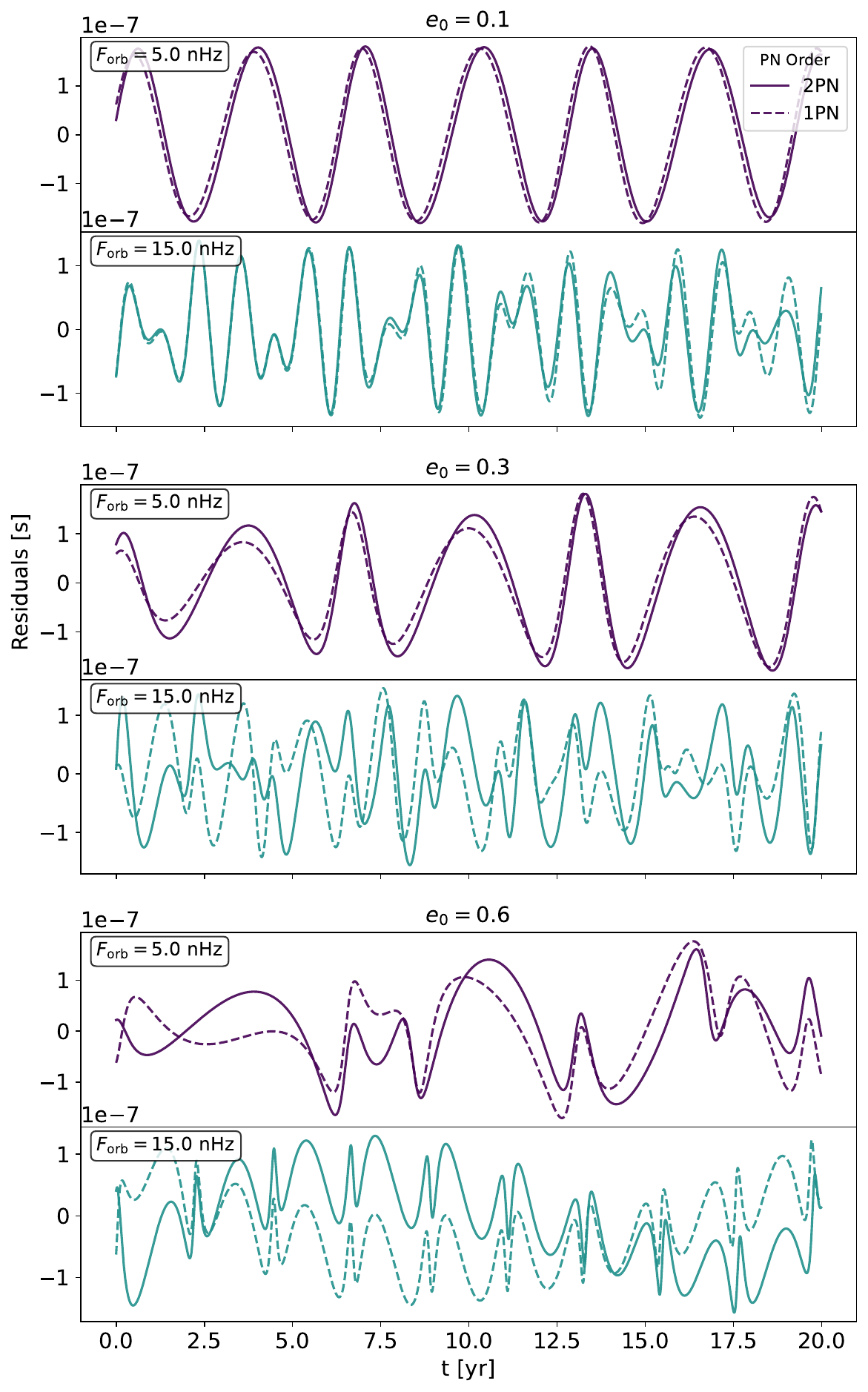}
\caption{\label{fig:res1/2pn} Complete residuals (ET+PT) computed assuming dynamics truncated at lowest PN order (dashed line) and high PN order (solid line) for different values of initial orbital frequencies: $F_{\text{orb}} = 5 \text{nHz}$ (purple) and $F_{\text{orb}} = 15 \text{nHz}$ (teal). The three panels correspond to different values of the initial eccentricity $e_0 = 0.1$ (up), $0.3$ (center), $0.6$ (bottom).}
\end{figure}
In Figure~\ref{fig:res1/2pn}, the residuals computed using the ODEs truncated at leading order (dashed line) are shown for visual comparison against the next-to-leading order ($2\text{PN}$) expansion of the ODEs (solid line). Already in the low-frequency ($F_{\text{orb}} = 5 \text{ nHz}$) and low-eccentricity ($e_0 = 0.1$) regime (purple in the upper panel), the effects of the $2\text{PN}$ expansion are slightly noticeable. These differences become more evident for mildly eccentric systems ($e_0 = 0.3$, central panels). Eventually, in the high-frequency ($F_{\text{orb}} = 15 \text{ nHz}$) and high-eccentricity ($e_0 = 0.6$) regime (teal in the bottom panel), the two waveforms completely disagree.

We want to emphasize an important caveat.
Although the 2PN terms are small, one might expect that ODEs truncated at the 1PN order could still reproduce the signal at the cost of introducing a parameter bias; in other words, the model with only 1PN dynamics could be effectual \cite{PhysRevD.67.064028}.
However, a PTA observes an array of pulsars, each with its own $\tau_{\alpha}$, effectively providing snapshots of the GW signal at different times.
A large array with a wide range of $\tau_{\alpha}$ values allows us to probe different segments of the GW signal, thus increasing the sensitivity to 2PN effects.
As discussed previously, the periapse precession already breaks the degeneracy between the individual component masses (since it depends on the total mass rather than the chirp mass).
The ability to measure 2PN terms, which depend explicitly on the mass ratio, will further improve the determination of individual masses. The necessity of including 2PN-order terms to correctly compute the pulsar term is the second important result of this paper.

\subsection{Residuals in frequency domain}\label{sec:fourier_decomposition}
Due to the presence of $({1+e\cos\xi})$ in denominator of residuals in Eq.~\eqref{eq:r+,x}, we lost a nice feature of arranging the residuals as harmonics of $k\xi + m\gamma$.  However we still can present our solution in harmonics of the mean anomaly $\psi_r$, see \autoref{fig:orbit}, which is equivalent to the Fourier decomposition. Indeed, 
following \cite{PhysRev.131.435}, we introduce mean anomaly $\psi_r = u - e\sin u$, where $u$ is eccentric anomaly,
$\psi_r = \omega t \equiv \omega_r t$ and $\omega = 2\pi F$ corresponds to Fourier frequency. We decompose the right-hand side of \eqref{eq:h+,x} in harmonics of $\psi_r$ using  
\begin{widetext}
\begin{subequations}\label{eq:bessel41}
  \begin{align}
    \frac{(e + 2\cos \xi) \sin\xi}{1+e \cos\xi} &= \frac{1}{\sqrt{1-e^2}}\left( \sum_{k=1}^{\infty} \left[\frac{2}{k} J_k(ke) - \frac{4}{e^2 k} J_k (ke) +\frac{2(1-e^2)}{e} \left(J_{k-1}(ke)-J_{k+1}(ke)\right)\right]\sin({k \psi_r})\right) \notag\\
    & = \sum_k a_k \sin({k \psi_r}) \label{subeqn:r_a_f}\\
    \frac{\cos 2\xi + e \cos \xi}{1+e \cos\xi} &= \sum_{k=1}^{\infty} \left[ \frac{4(1-e^2)}{e^2} J_k(ke) - \frac{2}{ke}\left(J_{k-1}(ke)-J_{k+1}(ke) \right)\right]\cos (k \psi_r) = \sum_k b_k \cos({k \psi_r})
    \label{subeqn:r_b_f}\\
    \frac{e \sin \xi}{1+e \cos\xi} &= \frac{e}{\sqrt{1-e^2}}\sum_{k=1}^{\infty}\frac{1}{k}\left( J_{k-1}(ke) + J_{k+1}(ke)\right) \sin({k \psi_r}) =  \sum_k c_k \sin({k \psi_r})
    \label{subeqn:r_c_f}
  \end{align}
\end{subequations}
\end{widetext}
Further details on the derivation of these expressions are provided in Appendix~\ref{sec:fourier}. We need to decompose the residuals into monotonic harmonics, taking into account the presence of two generally non-commensurate frequencies: the mean orbital frequency $\omega_r$ and the periapsis precession rate $\dot{\gamma}$. The first corresponds to the mean orbital motion, while the second represents the precession frequency.
In deriving Eq.~\eqref{eq:bessel41}, we have assumed purely conservative dynamics, and subsequently restored the time dependence on the right-hand side of Eq.~\eqref{eq:bessel41}.
The residuals can then be written as 

\begin{subequations}\label{eq:harms}
  \begin{align}
 r_{+} &= -\frac{iA}{2}\left[ 
    (1+\cos^2\iota) \sum_{k\ge 1} \left( G_k e^{i(k\psi_r +2\gamma)} + H_k e^{i(k\psi_r - 2\gamma)} \right) \right.\nonumber \\
   & \left.+  \sin^2{\iota} \sum_{k\ge 1}  c_k e^{i(k\psi_r)} \right]  + c.c.\\
 r_{\times} &= -A \cos{\iota}   \sum_{k\ge 1} \left( G_k e^{i(k\psi_r +2\gamma)} - H_k e^{i(k\psi_r - 2\gamma)} \right) + c.c.
  \end{align}
\end{subequations}
where {\it{c.c.}} denotes the complex conjugate of the preceding term, and 
$$
G_k(e) = a_k +b_k, \,\, H_k(e) = a_k - b_k.
$$

The expressions in Eq.~\eqref{eq:harms} are explicitly decomposed into harmonics of
$k\psi_r \pm \ell\gamma$, where $\ell = 0, \pm2$, but they are still written in the time domain. We use a stationary phase approximation applied to each harmonic to express residuals in the frequency domain.

We again consider the equal mass system with  $\mathcal{M}_c = 10^{9.2}M_{\odot}$,varying the initial eccentricity and mean orbital frequency $\omega_r = 2\pi F_0$. For simplicity, we restrict to a face-on binary configuration (the brightest case), $\cos{\iota} = 1$, where $r_{+} = r_{\times}$
\begin{figure}[h]
\centering
        \includegraphics[width=0.5\textwidth]{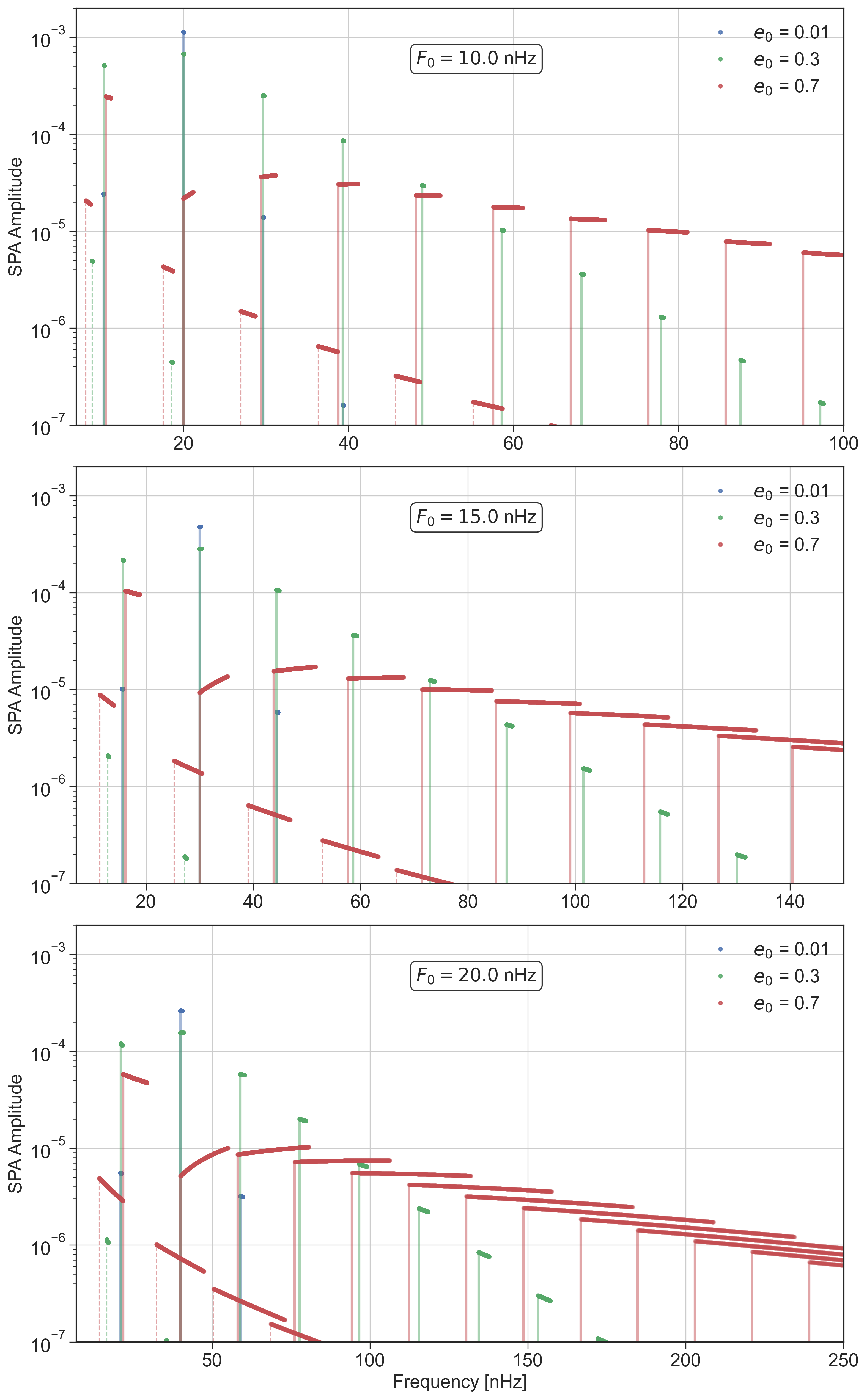}
    \caption{Harmonic decomposition of $r_{+}$ for a face-on system, showing the Earth-term only. Each panel corresponds to a different starting orbital frequency, and colours correspond to three initial eccentricities. Vertical lines assist in identifying individual harmonics: solid lines denote $k\psi_r + 2\gamma$, while dashed lines represent $k\psi_r - 2\gamma$.}
\label{fig:harms_ET}
\end{figure}
In \autoref{fig:harms_ET}, we show the harmonic content for binaries starting at orbital frequencies $F_0 = 10$, 15, and 20~nHz, with initial eccentricities $e_0 = 0.01, 0.3,$ and $0.7$. The nearly circular case ($e_0 = 0.01$) shows the expected dominance of the second orbital harmonic, while the first and third harmonics are suppressed by a factor $\sim 50$. For moderately eccentric binaries, the first three harmonics ($k=1,2,3$) dominate. At higher eccentricities, the spectrum becomes almost continuous due to a significant orbital evolution, with each harmonic spreading into a line. As noted in \cite{Taylor_2016}, the first orbital harmonic ($k=1$) dominates for large eccentricities. The strength of each harmonic of the eccentric binary is lower than that of the circular binary, but the power is spread over many Fourier bins and should be integrated. The thin vertical lines in the figure mark the harmonic positions, with solid lines for $k\psi_r + 2\gamma$ and dashed lines for $k\psi_r - 2\gamma$. The harmonics with negative precession are significantly lower in amplitude and most of them could be neglected. We observe an increase in orbital evolution as we move to more relativistic binaries (higher $F_0$), and even moderately eccentric systems show appreciable frequency drift.

\begin{figure}[h]
\centering
        \includegraphics[width=0.5\textwidth]{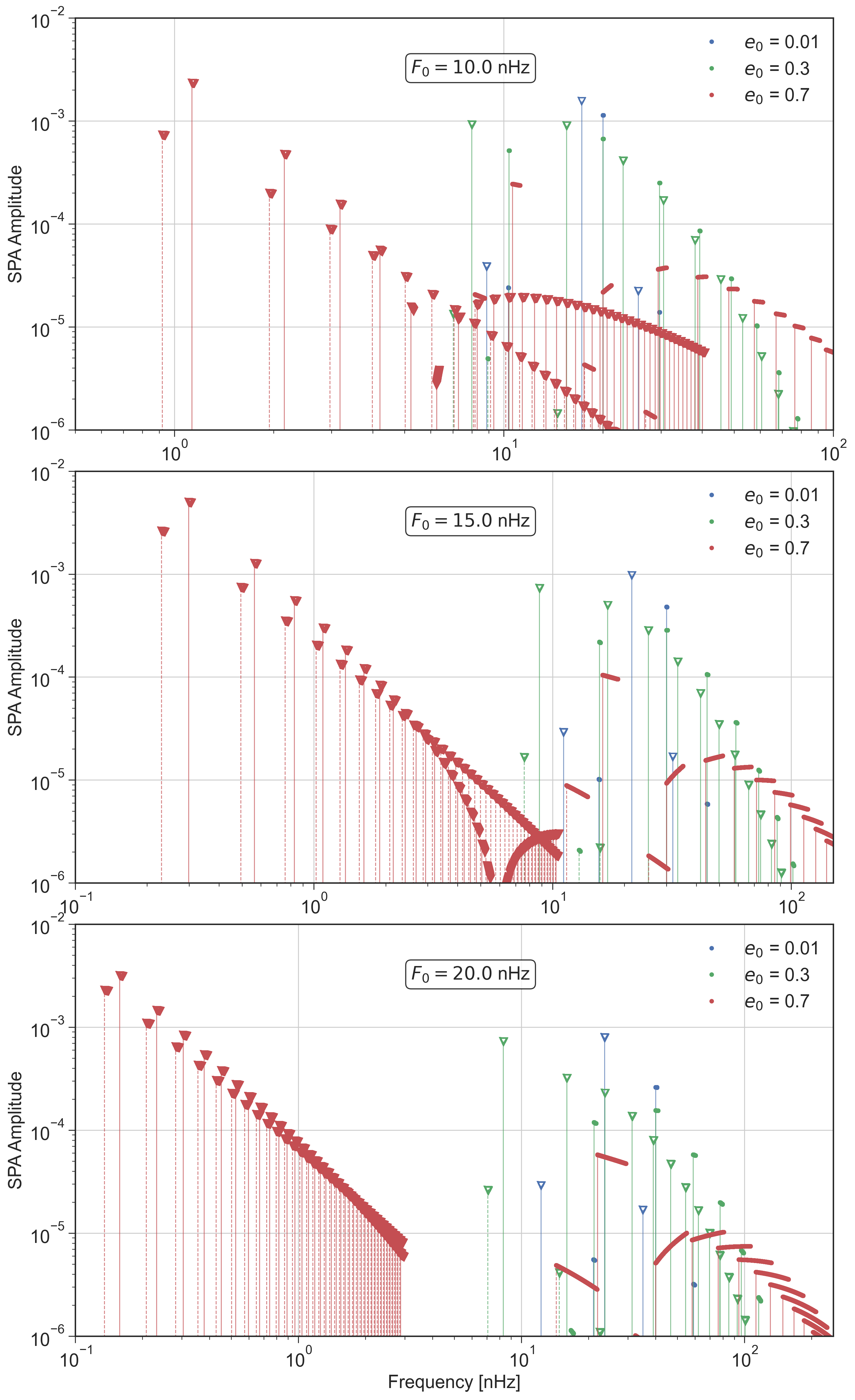}
  \caption{Extension of \autoref{fig:harms_ET} which now includes a contribution from the pulsar term shown as hollow triangles. We have included the frequency below the 
  lowest Fourier bin (1.6 nHz) to demonstrate the full spectrum.}
  \label{fig:harms_EPT}
\end{figure}

Next, we present the spectrum for the full $r_+$, which includes both the Earth and pulsar terms, in \autoref{fig:harms_EPT}.
We restrict the analysis to the first 50 harmonics.
The inclusion of the pulsar term introduces power at low frequencies, with its amplitude enhanced by the response factor $\propto 1/\omega_r$.
For circular binaries, we clearly identify two peaks corresponding to the second harmonic in Earth and pulsar terms.
However, the spectra of eccentric binaries show a much richer structure.
The $k=1$ harmonic of the pulsar term dominates the overall spectrum, although it is not always observable.
We extend the frequency range well below the first Fourier bin, $F_{\rm{min}} = 1/20_{\rm{yrs}} = 1.6~\text{nHz}$, to illustrate the full spectrum and to highlight that power from the pulsar-term harmonics can still leak into the observational frequency band.
Note that the plot spans four orders of magnitude in amplitude, and many of the displayed harmonics are too weak to be detected.
We emphasize the presence of very strong pulsar-term harmonics at low frequencies, which are incoherent across pulsars and could contribute to (or even dominate) the common red noise.
The mildly eccentric systems exhibit a mixture of Earth- and pulsar-term harmonics within the same frequency range.
The overall complexity of the spectrum underlines the importance of modelling the complete signal accurately, which includes the need for the post-Newtonian corrections to properly connect the orbital evolution between the Earth and pulsar terms.  The complex frequency structure of full GW induced signal in the PTA band is the third results of the paper.

\section{\label{sec:conclusion}Summary}

In this paper, we introduced a self-consistent model of the GW signal from eccentric  binaries in the PTA band.
This model is based on the EOB approach and assumes an adiabatic orbital evolution. We consider only non-spinning supermassive black holes and a generic mass ratio.

We derived the Hamiltonian equations governing the orbital dynamics and showed how they can be solved efficiently using an approximation valid across most of the parameter space.
Our analysis shows that the orbital evolution of eccentric binaries differs significantly from that of circular systems.
In particular, we highlight the importance of including 2PN terms when connecting the SMBHB orbit at the Earth and pulsar times.
The periapsis precession and 2PN corrections should help in breaking parameter degeneracies and enable the determination of the individual component masses; this will be explored in a separate publication.

Next, we obtained the GW signal by combining the dominant spherical harmonic modes,  $\ell = 2, m = 0,\pm2$, truncated at Newtonian order.  We transformed the GW strain into residuals in the time of arrival of radio pulses measured with the PTA experiment. The waveforms
are implemented in the time domain using \texttt{JAX}, the software will be available after the publication of this article. The use of \texttt{JAX} brings computational efficiency and enables automatic differentiation of the likelihood, which is essential for searches and machine learning applications.

Using the stationary phase approximation, we have investigated the spectrum of eccentric binaries. The use of a physically consistent GW model reveals a very reach harmonic structure in the eccentric timing residuals. In particular, we observe measurable harmonic evolution (especially for higher modes) and confirm the finding of \cite{Taylor_2016} that the first orbital harmonic dominates for mildly to highly eccentric binaries. We further demonstrated the importance of harmonics of the pulsar-term  and their mixture with the Earth-term across the same frequency range.  We emphasize that the complex spectral structure of residuals induced by eccentric GWs requires considering the full signal for its detection and correct interpretation of the observed data.

It is important to address the omission of spins in our analysis. Spin–orbit and spin–spin effects enter at 1.5PN and 2PN orders, respectively \cite{PhysRevD.47.R4183}. We expect that spin components aligned with the orbital angular momentum could produce a potentially measurable contribution when evolving the binary over $\tau_{\alpha}$ for the pulsar term. However, spins can likely be neglected over the observation span (20 years in this paper), unless the binary is at high frequency and high eccentricity (as in the top panel of \autoref{fig:evolution_eF}). At the same time, spin-induced additional (or missing) cycles may be covariant with the mass ratio \cite{PhysRevLett.109.081104}. Orbital precession caused by spin misalignment should be less important and could be accommodated by allowing for different inclinations for the Earth and pulsar terms.
These conjectures require a more careful treatment using an EOB Hamiltonian for spinning MBHBs, such as those described in \cite{PhysRevD.105.044035, Nagar_2021, Chiaramello_2020}. We note that different EOB implementations are expected to agree at the modest PN orders relevant for PTA.

Finally, we want to emphasize that the eccentric gravitational-wave model developed in this work is equally applicable to Galactic stellar-mass binaries (white dwarfs, neutron stars and black holes) that will be observed by LISA.

\begin{acknowledgments}
We acknowledge funding from the French National Research Agency
(grant ANR-21-CE31-0026, project MBH\_waves).
The authors also acknowledge support from the CNES for the exploration of
LISA science.
\end{acknowledgments}

\appendix

\section{Mapping EOB to ADM coordinates}\label{sec:ADM}
In this appendix, we establish the mapping between the EOB coordinates (subscript “E”) and the ADM coordinates (subscript “A”) adopted in \cite{Susobhanan:2022nzv}, accurate through first post-Newtonian (1PN) order.
Post-Newtonian calculations typically rely on the quasi-Keplerian (QK) parameterization, which provides an explicit solution to the two-body dynamics as a perturbative correction to Newtonian motion \cite{Damour_2004}.
This approach introduces several auxiliary variables: three “eccentricities” $(e_t, e_r, e_{\phi})$, a semi-major axis $a$, and three angular parameters — the true anomaly ($\nu$), eccentric anomaly ($u$), and mean anomaly ($l$) — in addition to the azimuthal orbital phase ($\phi$). The relative radial separation is expressed in terms of the eccentric anomaly $u$ as
\begin{equation}
r^{\text{A}} = a(1-e_r \cos u).
\end{equation}
To map our coordinate system to the ADM frame, we express the dynamics in terms of gauge-invariant quantities.
We expand all orbital elements as PN series in powers of $\epsilon \equiv -2E$, with coefficients depending on $\nu$ and $j \equiv -2Eh^2$, where $h \equiv J/M$ \cite{Arun_2008}, and $E$, $J$ are the orbital energy and angular momentum of the binary.
In many ADM-based treatments, including \cite{Susobhanan:2022nzv}, the dynamics is parameterized by the mean motion $n$ rather than the azimuthal frequency, with
$x_r = M n = M \omega_r$, and the periastron advance defined as $K = 1 + k$, yielding $x_{\phi} = M n (1 + k)$.
Since $n$ and $K$ are coordinate-invariant functions of $(\epsilon, j)$ \cite{Damour_1988},
\begin{equation}
n_{\text{E}} = n_{\text{A}}, \qquad K_{\text{E}} = K_{\text{A}}.
\end{equation}

To match quantities between ADM and EOB coordinates, we relate $x_r$ and $x_{\phi}$, which coincide at Newtonian order but differ starting at 1PN order \cite{Arun_2008}:
\begin{equation}\label{eq:xi_vs_x}
x_r(x_{\phi}) = \frac{x_{\phi}^{3/2}\left(1 - e_t^6 + 3 e_t^4 - 3 e_t^2 + (-3 e_t^4 + 6 e_t^2 - 3) x_{\phi}\right)}{(1 - e_t^2)^3}.
\end{equation}

Starting from the ADM energy expression \cite{Damour_2004} and substituting Eq.~\eqref{eq:xi_vs_x}, we find
\begin{equation}
\epsilon_{\text{A}} = -2E = x_r^{2/3}\left[1 + \frac{x_r^{2/3}}{12}(15 - \eta)\right].
\end{equation}
Replacing $x_{\phi}$ by the semilatus rectum $p$ using Eq.~\eqref{eq:x} (1PN truncation) yields
\begin{equation}\label{eq:epsilon}
\epsilon_{\text{A}} = \epsilon_{\text{E}} = \frac{1 - e^2}{p} + \frac{(e^2 - 1)^2(\nu - 3)}{4p^2} + O(p^{-3}).
\end{equation}
A similar procedure yields the mapping for the angular momentum.
In ADM coordinates \cite{Damour_2004}, we have
\begin{equation}
j_{\text{A}} = -2Eh^2 = (1 - e_t^2)\left[1 - \frac{x_r^{2/3}}{4(1 - e_t^2)}\big((17 - 7\nu)e_t^2 - 9 - \nu\big)\right].
\end{equation}
After converting $x_r \rightarrow x_{\phi}$ (Eq.~\eqref{eq:xi_vs_x}), $x_{\phi} \rightarrow p$ (Eq.~\eqref{eq:x}), and $e_t \rightarrow e$, with
\begin{subequations}
\begin{align}
    e_t &= \sqrt{1-j +\frac{\epsilon}{4} (-8+8 \nu -j (7 \nu-17))} \\
    & = e-\frac{e \left(e^2-1\right) (\nu-3)}{p}+O\left(\left(\frac{1}{p}\right)^2\right),
\end{align}
\end{subequations}
we obtain
\begin{equation}\label{eq:j}
    j_{\text{A}} = j_{\text{E}} = \left(1-e^2\right)\left(1+\frac{9+\nu + e^2(7-\nu)}{4p}\right)+O\left(\frac{1}{p^2}\right).
\end{equation}

Next, we compare the equations of motion for $\dot{r}$ and $\dot{\phi}$ from Eqs.~\eqref{eq:xidot} and \eqref{eq:EOMs_cons}, converting them to ADM coordinates using
\begin{equation}
    \delta_r = -M + \nu \left(\frac{3}{2}r\dot{r}^2 + \frac{1}{2}r^3\dot{\phi}^2 - \frac{M}{2}\right),
\end{equation}
\begin{equation}
    \delta_{\dot{r}} = \nu \dot{r} \left(\frac{5}{2} v^2 - \dot{r}^2 - 3\frac{M}{r}\right),
\end{equation}
where $v^2 = \dot{r}^2 + r^2\dot{\phi}^2$ and $r$ is given by Eq.~\eqref{eq:rfuncxi} \cite{Hinderer_2017}.
This gives
\begin{subequations}
    \begin{align}
        r_A &= r_E + \delta_r, \\
        \dot{r}_A &= \dot{r}_E + \delta_{\dot{r}},
    \end{align}
\end{subequations}
followed by variable changes $x_{\phi} \rightarrow x_r$ and $e \rightarrow e_t$.

In many formulations, the dynamics is expressed in terms of the eccentric anomaly $u$ rather than the true anomaly $\xi$.
The mapping between the two is
\begin{equation}\label{eq:cos_u}
\cos u = \frac{1}{a_r e_r}\left[a_r - \left(\frac{Mp}{e\cos\xi + 1} + \delta_r\right)\right],
\end{equation}
from which $\sin u$ can also be derived and both expanded in powers of $1/p$ up to first order.
The major axis $a_r$ and the auxiliary eccentricity $e_r$ are functions of $\epsilon$ and $j$:
\begin{equation}
    e_r = \sqrt{1-j+\frac{\epsilon}{4}(5 j (\nu-3)-4 \nu+24)},
\end{equation}
\begin{equation}
    a_r =\frac{M}{\epsilon} \left(1+\frac{1}{4} \epsilon (\nu-7)+\frac{1}{16} \epsilon^2 \left(\frac{44 \nu-68}{j}+\nu^2+10 \nu+1\right)\right),
\end{equation}
then substituted using Eqs.~\eqref{eq:epsilon} and \eqref{eq:j} expanded in $1/p$.

\section{Fourier decomposition of Keplerian motion}\label{sec:fourier}

The Fourier analysis of Kepler's equations 
\begin{equation}
    \psi_r = u - e\sin u
\end{equation}
is given by the Kapteyn series:
\begin{equation}
    u = \psi_r + 2\sum_{k=1}^{\infty} \frac{J_k(k e)}{k} \sin({k \psi_r})
\end{equation}
from which it is possible to obtain the following combinations of sine/cosine of $u,\xi$:
\begin{widetext}
\begin{subequations}\label{eq:bessel}
  \begin{align}
    &\sin u =\frac{2}{e}  \sum_{k=1}^{\infty} \frac{J_k(k e)}{k} \sin({k \psi_r}) \label{subeqn:sinu}\\
    &\cos\xi \sin u =\sum_{k=1}^{\infty}\left(-\frac{2}{e^2} \frac{J_k(k e)}{k} + \frac{2(1-e^2)}{e}  \frac{J'_k(k e)}{k}\right)\sin({k \psi_r}) \label{subeqn:cosxisinu}\\
    & \sin u \sin \xi = \sqrt{1-e^2}\sum_{k =1}^{\infty}\bigg( \frac{1}{2} - \frac{2(1-e^2)}{e^2}  J_k(ke)
    +\frac{2}{e}\frac{J'_k(ke)}{k}\bigg)\cos(k\psi_r)\label{subeqn:sinusinxi}
  \end{align}
\end{subequations}
\end{widetext}
Expressing terms in the residuals in the Fourier domain requires a partial change of variable from $\xi$ to $u$ according to
\begin{subequations}\label{eq:r_abc}
  \begin{align}
    &\frac{(e + 2\cos \xi) \sin\xi}{1+e \cos\xi} = \frac{1}{\sqrt{1-e^2}}\left(e \sin u + 2 \cos\xi\sin u\right), \label{subeqn:r_a}\\
    &\frac{\cos 2\xi + e \cos \xi}{1+e \cos\xi} = 1-\frac{2\sin u \sin \xi}{\sqrt{1-e^2}}, \label{subeqn:r_b}\\
    &\frac{e \sin\xi}{1+e\cos\xi} = \frac{e\sin u}{\sqrt{1-e^2}}. \label{subeqn:r_c}
  \end{align}
\end{subequations}
We can now use expressions given in \eqref{eq:bessel} and apply the recursive formula for derivatives of Bessel's functions: $J'_k(ke)= \frac{1}{2} \left( J_{k-1}(ke) - J_{k+1}(ke)\right)$.

\nocite{*}

\bibliography{RefsSara}

\end{document}